\documentclass[12pt]{article}
\usepackage{amsfonts}
\newtheorem{theo}{Theorem}[section]
\newtheorem{lemma}{Lemma}[section]
\newtheorem{defi}{Definition}[section]

\newcommand{\proofend}{\raisebox{1.3mm}{\fbox{%
 \begin{minipage}[b][0cm][b]{0cm}\end{minipage}}}}
\newenvironment{proof}{\noindent{\it Proof:} }{\mbox{}\hfill
 \proofend\\\mbox{}}

\newenvironment{rem}{\noindent{\it Remark:} }{\medskip}

\newcommand{\Ab}{\overline{{\cal A}}}
\newcommand{\Gb}{\overline{{\cal G}}}

\newcommand{\Ub}{\overline{{\cal U}}}
\newcommand{\AbR}{\overline{{\cal A}}_B}
\newcommand{\GbR}{\overline{{\cal G}}_B}

\newcommand{\UbR}{\overline{{\cal U}}_B}
\newcommand{\AbS}{\overline{{\cal A}}_{\Sigma}}
\newcommand{\GbS}{\overline{{\cal G}}_{\Sigma}}

\newcommand{\TAb}{\overline{{\cal T}}}
\newcommand{\PAb}{\overline{{\cal P}}}
\newcommand{\Ag}{{\cal A}_{\gamma}}
\newcommand{\Gg}{{\cal G}_{\gamma}}
\newcommand{\Ug}{{\cal U}_{\gamma}}
\newcommand{\Lie}{{\cal L}}
\newcommand{\pgg}{p_{\gamma\gamma^{\prime}}}
\newcommand{\Phip}{\Phi^{\prime}}
\newcommand{\lp}{\lambda^{\prime}}
\newcommand{\rstar}{\overline{r}_{[\lambda]}^{\star}}
\newcommand{\rnstar}{\overline{r}_{[\lambda_n]}^{\star}}
\newcommand{\sigl}{\sigma_{[\lambda]}}
\newcommand{\sigln}{\sigma_{[\lambda_n]}}
\newcommand{\kt}{\vartheta}
\newcommand{\kp}{\varphi}
\newcommand{\Eg}{E(\gamma)}
\newcommand{\Vg}{V(\gamma)}

\newcommand{\md}{\mathchoice%
  {\mbox{\rm d}}%
  {\mbox{\rm d}}%
  {\mbox{\scriptsize\rm d}}%
  {\mbox{\tiny\rm d}}}
\newcommand{\td}[2][]{\mathchoice%
  {\frac{\mbox{\small\rm d}#1}{\mbox{\small\rm d}#2}}%
  {\frac{\mbox{\scriptsize\rm d}#1}{\mbox{\scriptsize\rm d}#2}}%
  {\frac{\mbox{\tiny\rm d}#1}{\mbox{\tiny\rm d}#2}}%
  {\frac{\mbox{\tiny\rm d}#1}{\mbox{\tiny\rm d}#2}}}
\newcommand{\pd}[2][]{\frac{\partial #1}{\partial #2}}
\newcommand{\fd}[2][]{\frac{\delta #1}{\delta #2}}
\newcommand{\Aut}{\mbox{\rm Aut}\,}
\newcommand{\Ad}{\mbox{\rm Ad}\,}
\newcommand{\ad}{\mbox{\rm ad}\,}
\newcommand{\Hom}{\mbox{\rm Hom}\,}
\newcommand{\Ima}{\mathchoice%
  {\mbox{\rm Im}\,}%
  {\mbox{\rm Im}\,}%
  {\mbox{\scriptsize\rm Im}\,}%
  {\mbox{\tiny\rm Im}\,}}
\newcommand{\id}{\mbox{\rm id}\,}
\newcommand{\diag}{\mbox{\rm diag}\,}
\newcommand{\Kern}{\mbox{\rm ker}\,}
\newcommand{\tr}{\mbox{\rm tr}\,}

\newcommand{\DirB}[3][\mkern-5mu]{\ensuremath{%
 \left\langle#2\left|#1\right|#3\right\rangle}}

\newcommand{\ket}[1]{\ensuremath{\left|#1\right\rangle}}

\newcommand*{\R}{{\mathbb R}}
\newcommand*{\N}{{\mathbb N}}
\newcommand*{\Z}{{\mathbb Z}}

\newcommand*{\C}{{\mathbb C}}

\begin{document}\begin{titlepage}\begin{flushright} PITHA 99/23 \\
hep-th/9907042
\end{flushright} \vspace{1cm}\begin{center}
{\LARGE Quantum Symmetry Reduction \vspace{0.2cm} \\for
Diffeomorphism Invariant Theories \vspace{0.3cm} \\of Connections}
\vspace{1cm}
\\ {\large M.~Bojowald\footnote{e-mail address: {\tt
bojowald@physik.rwth-aachen.de}} ~and H.A.\
Kastrup\footnote{e-mail address: {\tt
kastrup@physik.rwth-aachen.de}}} \vspace{0.5cm}
\\ {\normalsize Institute for Theoretical
Physics\vspace{0.1cm} \\ RWTH Aachen, D-52056 Aachen, Germany}
\end{center} \vspace{0.5cm}

\begin{abstract}
Given a symmetry group acting on a principal fibre bundle,
symmetric states of the quantum theory of a diffeomorphism
invariant theory of connections on this fibre bundle are defined.
These symmetric states, equipped with a scalar product derived
from the Ashtekar-Lewandowski measure for loop quantum gravity,
form a Hilbert space of their own. Restriction to this Hilbert
space yields a quantum symmetry reduction procedure in the
framework of spin network states the structure of which is
analyzed in detail.

Three illustrating examples are discussed: Reduction of $3+1$ to
$2+1$ dimensional quantum gravity, spherically symmetric quantum
electromagnetism and spherically symmetric quantum gravity.
\end{abstract}
\end{titlepage}
\section{Introduction}

During the last years there have been many very active and
partially successful attempts to quantize gravitational systems:
One very ambitious approach is string theory (see, e.g.\ Ref.\
\cite{po}) which tries to cover the quantum theory of all
interactions ``at once''. Another one is loop quantum gravity (see
the reviews \cite{ro1}) which uses a (non-perturbative) canonical
framework (Cauchy surfaces, canonical coordinates and momenta,
constraints etc.) for diffeomorphism invariant gravity formulated
in terms of connection variables, without or with couplings to
matter.

 Among its achievements loop quantum gravity counts the following
 ones: The uncovering of a discrete structure of space
\cite{AreaVol,Area,Vol2,thi3}, a derivation  of the
Bekenstein-Hawking formula for the black hole entropy in terms of
microscopic degrees of freedom \cite{R:LoopEntro1,R:LoopEntro2,
K:LoopEntro1,K:LoopEntro2,ABCK:LoopEntro} and a regularization and
partial solution of  the constraints of general relativity
\cite{ALMMT,AnoFree,QSDI,QSDII}. All these results rely heavily on
the fundamental assumption that holonomies (Wilson loops) of the
$SU(2)$-gauge connection of general relativity in the Ashtekar
formulation with {\em real} connections \cite{AshVar,AshVarReell}
become densely defined operators in the quantum theory.

In order to test the basic assumptions of that approach and in
order to check whether the theory can have the correct classical
limit, mini- and midi-superspace models \cite{midisup} are very
appropriate because their (strongly) reduced number of degrees of
freedom make them more transparent and sometimes even solvable.

These models are usually obtained by a classical symmetry
reduction of the full theory and they are selected in order to
facilitate quantization while keeping some of the basic problems
faced in the quantization of general relativity. An example is the
reduction of the $3+1$ dimensional theory to a $2+1$ dimensional
one. Here a simplification occurs because $2+1$ dimensional
gravity has only finitely many physical degrees of freedom and it
is exactly soluble \cite{WittenZweiPlusEins}. This model has
already been investigated in the loop quantization approach
\cite{QSDIV} starting from a classical $2+1$ dimensional theory of
gravity in terms of Ashtekar variables.

Our strategy will be different: We want to isolate symmetric
states of the already loop quantized $3+1$ dimensional full
theory. These states have to be exactly symmetric, not only
symmetric at large distances compared to the Planck scale.
Constraining the space of physical states to these symmetric ones
amounts to a quantum symmetry reduction.

Because the solutions to all the constraints of loop quantum
gravity are not known we will carry out this symmetry reduction
procedure on an auxiliary Hilbert space. Therefore we will have to
regularize and to solve the reduced constraints on our spaces of
symmetric states. As always with reduced models, the hope is that
this regularization and the search for solutions can be done with
more ease. At the same time one hopes that the model under
consideration can lead to new insights in the quantization of the
full theory.

 Our main motive for the analysis of the  present
paper has its origin in our interest in the reduction to spherical
symmetry. The classically reduced Schwarzschild system has been
quantized using Ashtekar variables (but not using loop
quantization techniques) in Refs.\ \cite{thika,kathi}. A
corresponding analysis has been performed for the
Reissner-Nordstr{\o}m model \cite{thi2}. These models are of
physical interest because they are related to vacuum black holes.
Similar to $2+1$ dimensional gravity they have only finitely many
physical degrees of freedom, and it is an interesting question how
this reduction of infinitely many degrees of freedom of the full
theory takes place in a loop quantization using spin network
states.

Another motivation is to possibly find a way in order to calculate
the degeneracy of energy levels. The levels are not degenerate in
the quantization of the classically reduced theory; however, this
approach makes use only of smooth fields whereas loop quantum
gravity  relates  the black hole entropy  to distributional
configurations (generalized functions). Also, non-spherical
fluctuations should not be ignored.  The degeneracy plays a
crucial role in the calculation \cite{ka1,ka2} of black hole
entropy using a canonical partition function approach. The entropy
obtained in this way is proportional to the horizon area, but the
constant of proportionality  is only known to be $O(1)$, so that a
quantitative comparison with the semiclassical $A/4\;$-law is not
immediate. A similar problem arises in the loop quantum gravity
calculation \cite{ABCK:LoopEntro} because of the so-called Immirzi
parameter \cite{imm}. A comparison of the entropy of the full
theory with that of the spherically symmetric one may shed some
light on the origin of the degeneracies because degenerate states
of the symmetric model may become non-degenerate for the
non-symmetic one.

Another possible application is to study cosmological models
within loop quantum gravity \cite{bo1}.

In order to attack these physical questions we have to define
symmetric states in the Hilbert space of loop quantum gravity and
have to  determine their properties. This will be done in the
present paper in a more general setting: We will investigate the
case of a Lie symmetry group $S$ acting on a principal fibre
bundle with a compact Lie structure group $G$.

 Before going to this general case let us dwell briefly on spherically
symmetric quantum gravity:  The states of the (non-spherically
symmetric) full theory are given in terms of a polymer-like
structure called spin network, lying in the spacelike section of
space-time used to carry out a canonical quantization.  We now
have to face the problem how to establish a symmetry of a discrete
structure under a continuous symmetry group. A possible approach
is suggested by the well-known solution of the diffeomorphism
constraint using group averaging methods \cite{ALMMT}. However,
this cannot lead to the desired result here: We would have to
average not only over rotated subgraphs of the graph underlying a
given spin network which is to be averaged, but it would be
necessary to average over all rotations of an edge while keeping
the other edges fixed because these give the same holonomy when
evaluated in an invariant connection. Otherwise the holonomy of an
edge as a multiplication operator would not commute with
rotations. Therefore some averaging of parts of a given graph has
to be done, which, however, runs into problems when gauge
invariance has to be imposed. In some sense the rotation group is
too rigid as compared to the diffeomorphism group: It acts only
simply transitively on its orbits, whereas the diffeomorphism
group acts $k$-transitively for any $k \in \N$, i.e.\ any two
given sets of $k$ different points can be mapped one onto another
by a single diffeomorphism.

A lesson, however, can nevertheless be drawn from group averaging:
Symmetric states have to be  distributions (generalized functions)
on the function space over the quantum configuration space. This
can be understood from the following intuitive picture:
Constraining the support of a spin network function to only
symmetric connections yields a singular distribution. This
observation will guide us in our definition of symmetric states.

In order to achieve this we will use the theory of invariant
connections on symmetric principal fibre bundles
\cite{KobNom,Harnad,Brodbeck}. The essential properties and
results will be recalled in the next section, together with some
mathematical techniques of loop quantization. That section will
also serve to fix our notation.

Section~\ref{s:SymmState} deals with the definition of symmetric
states and the analysis of their properties; it contains entirely
new material. Our main result is Theorem \ref{ident}, by means of
which we can identify spaces of symmetric states with certain spin
network spaces.

 In Section~\ref{s:Examples} we will give some
examples: Quantum symmetry reduction to $2+1$ dimensional gravity
and spherically symmetric electromagnetism as well as gravitation.
These examples are intended to illustrate the ``kinematical''
framework of the symmetry reduction proposed here. Solving the
corresponding constraints is another task.

 As to $2+1$ dimensional
gravity we shall show how our approach leads to the same results
as that of Thiemann \cite{QSDIV}. Spherically symmetric
diffeomorphism invariant electromagnetism will be treated for a
vanishing gravitational field only. It is nevertheless an
interesting example for the symmetry reduction of a diffeomorphism
invariant system, and it illustrates the reduction of degrees of
freedom to finitely many ones and also the classifying role of the
magnetic charge.

The discussion of spherically symmetric gravity is mainly
restricted to kinematical aspects: the symmetry reduction is
implemented, the Gau{\ss} constraint solved in the context of the
appropriate spin networks and the solution of the diffeomorphism
constraint by group averaging  indicated. The problem of dealing
with the more difficult Hamiltonian constraint (definition as a
well-defined operator on spin-network states and the solution of
the constraint) will be discussed elsewhere \cite{bo3}.

An interesting application of our results to the spectrum of the
area operator acting in the spherically symmetric sector of loop
quantum gravity will be published in a separate note \cite{bo4}.
\section{Preparations}

In this section we will recall some facts concerning loop
quantization techniques and the theory of invariant connections on
symmetric principal fibre bundles which will be used in the
following sections.

\subsection{Spin Networks with Higgs Field Vertices}\label{s:spinnet}

Let $G$ be a compact Lie group, $\Sigma$ an analytic manifold and
$P(\Sigma,G,\pi)$ a principal fibre bundle over the base manifold
$\Sigma$ with structure group $G$. The affine space of connections
on this fibre bundle will be denoted as ${\cal A}$, and the
(local) gauge group as ${\cal G}$. Investigating invariant
connections will lead us to the use of Higgs fields, which are
sections of the adjoint bundle of $P$ (the associated vector
bundle employing the adjoint representation). The space of all
smooth Higgs fields will be called ${\cal U}$. These three
function spaces will have to be extended in the course of
quantization. Their treatment is given in Ref.\ \cite{ALMMT} for
${\cal A}$ and ${\cal G}$, and in Ref.\ \cite{FermionHiggs} for
${\cal U}$. In the following we will combine these procedures.

Let ${\cal T}$ be the parallel transport algebra generated by
elements of the matrix-valued parallel transporters associated
with a fundamental representation of $G$ along all piecewise
analytic paths in $\Sigma$, subject to appropriate relations
ensuring the correct matrix multiplication under composition of
paths and taking care of the group properties of $G$
(incorporating Mandelstam identities).

Similarly, let ${\cal P}$ be the point holonomy algebra generated
by elements of matrices, including the identity, in a fundamental
representation of $G$ of point holonomies. According to Ref.\
\cite{FermionHiggs} a point holonomy is a function on the space
${\cal U}$ of classical Higgs fields obtained by exponentiating
the value of a Higgs field at a given point.

 The multiplication within the algebras is
multiplication of $\C$-valued functions on ${\cal A}$ and ${\cal
U}$, respectively. Let $\TAb$ and $\PAb$ be their completions in
the sup norm.  These two algebras are abelian $C^{\star}$-algebras
with identity. We can build the product algebra $\TAb\otimes\PAb$
 which is the completed tensor product space of the underlying
vector spaces with pointwise multiplication. This, too, is an
abelian $C^{\star}$-algebra with identity and we can use
Gel'fand-Neumark theory (see, e.g.\ Refs. \cite{funct}) to obtain
the following isometry of $C^{\star}$-algebras:
\begin{equation}
 \TAb\otimes\PAb\cong C(\Ab)\otimes C(\Ub)\cong C(\Ab\times\Ub).
\end{equation}
$\Ab$ and $\Ub$ are the Gel'fand spectra of the respective
algebras consisting of all continuous $\star$-homomor\-phisms of
the algebras to $\C$. The isometry is given by the Gel'fand
transform $\hat{\:}\colon\TAb\to C(\Ab)$ defined below, and
similarly for $\Ub$. $\Ab$ and $\Ub$ are extensions of the spaces
${\cal A}$ and ${\cal U}$ which are densely embedded in the
Gel'fand topology. This topology is uniquely defined by the
following two conditions: $\Ab$ be compact and
$\hat{T}\colon\Ab\to\C,A\mapsto\hat{T}(A)=A(T)$ be continuous for
all $T\in\TAb$.  The compact Hausdorff space $\Ab\times\Ub$ will
serve as quantum configuration space.

There is, however, an alternative construction of these spaces
which is better suited for calculations. Here the extensions are
constructed as certain projective limit spaces.  The partially
ordered directed set used to define these limits is the set
$\Gamma$ of all piecewise analytic graphs $\gamma$ in $\Sigma$.
The projective families $(\Ag,\pgg)$ and $(\Gg,\pgg)=(\Ug,\pgg)$
can be found in Refs. \cite{AAla,ALMMT}. $\Ag$ is the space of all
functions which assign elements of the group $G$ to the edges of
$\gamma$, and which obey certain relations ensuring the correct
behavior under inversion and composition of edges. The elements of
$\Gg$ and $\Ug$ assign group elements to the vertices of $\gamma$.
The projections $\pgg\colon{\cal A}_{\gamma^{\prime}}\to\Ag$,
$\gamma\subset\gamma^{\prime}$ restrict the domain of definition
of the connections in ${\cal A}_{\gamma^{\prime}}$ to the edges of
$\gamma$, and similarly for $\Gg$ and $\Ug$. $\Gg$ acts on both
$\Ag$ and $\Ug$ by usual gauge transformations.  The projective
families define the projective limits $\Ab$, $\Gb$ and $\Ub$,
where now $\Gb$ acts on $\Ab$ and $\Ub$. These projective limit
spaces are identical to the Gel'fand spectra constructed above,
and their topology induced from the Tychonov topology is
equivalent to the Gel'fand topology. This can be seen from the
fact that $\Ab$ is --~as the projective limit of compact spaces~--
compact, and that the maps $\hat{T}$ are continuous.  In the proof
of Theorem~\ref{sigl} we will make use of the equivalence of
Gel'fand and Tychonov topologies.

Our quantum configuration space now is the space $\Ab\times\Ub$
and the auxiliary Hilbert space will consist of functions on this
space. An important class of such functions is that of cylindrical
functions which depend on the connection and Higgs field only via
a finite number of edges and vertices in $\Sigma$. A function $f$,
cylindrical with respect to the underlying graph $\gamma$, can be
written as
\begin{equation}
 f(A,U)=f_{\gamma}(A(e_1),\ldots,A(e_n),U(v_1),\ldots,U(v_m))
\end{equation}
where $e_1,\ldots,e_n$ are the edges of $\gamma$, and
$v_1,\ldots,v_m$ its vertices. The functions $f_{\gamma}$
representing a cylindrical function $f$ have to obey certain
consistency conditions.  The auxiliary Hilbert space is
$L_2(\Ab\times\Ub,\md\mu_{AL})$ obtained from the space of
cylindrical functions by completion with respect to the
Ashtekar-Lewandowski measure $\md\mu_{AL}$, which is, on a
cylindrical subspace, the finite product of Haar measure on $G$
for each edge and vertex of the respective graph $\gamma$.

An orthogonal basis is given by the set of spin network functions
with Higgs field vertices. These are cylindrical functions given
by a graph $\gamma$ together with a labeling $j$ of its edges, and
labelings $j^{\prime}$, $j^{\prime\prime}$ of its vertices with
equivalence classes of irreducible representations of $G$, and a
third labeling $C$ of the vertices with certain intertwining
operators.  Given a vertex $v$, $C_v$ is given by an intertwining
operator from the tensor product of the representations $j_e$
labeling incoming edges $e$ and the representation $j_v^{\prime}$
to the tensor product of the representations $j_e$ labeling
outgoing edges and the representations $j_v^{\prime}$ and
$j_v^{\prime\prime}$. The value of a spin network function on an
element $(A,U)\in\Ab\times\Ub$ is found by taking for each edge
$e\subset\gamma$ the element $A(e)$ in the representation $j_e$
and for each vertex $v\in\gamma$ the element $U(v)$ in the
representation $j_v^{\prime}$, and then contracting these matrices
according to the intertwining operators in the vertices. The
resulting function will transform according to the representation
$j_v^{\prime\prime}$ in each vertex $v$. In particular, the spin
network function will be gauge invariant if all the
representations $j_v^{\prime\prime}$ are trivial.

\subsection{Invariant Connections on \\ Symmetric Principal Fibre Bundles}
\label{s:invconn}

It is well known \cite{KobNom,Harnad,Brodbeck} that an invariant
connection on a manifold $\Sigma$ can be decomposed into a reduced
connection of a reduced gauge group on a submanifold
$B\subset\Sigma$ plus some scalar fields on $B$ acted on by a
group action determined by the symmetry reduction (a
representation of the reduced structure group). The multiplet of
scalar fields will be called ``Higgs field'' in the following. It
arises because in general an invariant connection is not
manifestly invariant, but only invariant up to gauge
transformations.

E.g.\ the authors of \cite{CorderoTeit,Cordero} make an ansatz for
a spherically symmetric connection using the fact that a symmetry
transformation can be compensated by a gauge transformation if the
Lie algebra of the structure group contains a $su(2)$ subalgebra.

In this paper we will use a more general and more systematic
approach which yields a complete classification of invariant
connections on symmetric principal fibre bundles. It can be found
in Refs.\ \cite{KobNom,Harnad,Brodbeck}, and its main elements
will be recalled in the present subsection.  The method has the
following advantages:

\begin{itemize}
 \item The structure of the reduction of the gauge group and the
   appearance of Higgs fields becomes clearer.
 \item All partial gauge
   fixings (selections of a certain
   homomorphism $\lambda\in[\lambda]$ defined
   below) can be treated on the same footing (and eventually be
   relaxed), whereas the ansatz of Refs.\
   \cite{CorderoTeit,Cordero} amounts to selecting
   one special $\lambda$, i.e. a partial gauge fixing.
 \item A
   possible topological charge given by gauge inequivalent actions of
   the symmetry group in the fibres can be taken into account. This is
   excluded by the ansatz of \cite{CorderoTeit,Cordero} from the
   outset by using  trivial bundles and a fixed action of the
   symmetry group on the bundle only.
\end{itemize}

Whereas the first two points will be essential for constructing
symmetry reductions in the spin network context
(Section~\ref{s:SymmState}), the last point is needed for
generality and allows to describe e.g. a magnetic charge.

Now let $S<\Aut(P)$ be a Lie symmetry subgroup of bundle
automorphisms acting on the principal fibre bundle
$P(\Sigma,G,\pi)$ defined above. Using the projection $\pi\colon
P\to\Sigma$ we get a symmetry operation of $S$ on $\Sigma$. For
simplicity we will assume that all orbits of $S$ are of the same
type; if necessary we will have to decompose the base manifold in
several orbit bundles $\Sigma_{(F)}\subset\Sigma$, where $F=S_x$
is the isotropy subgroup of $S$ consisting of elements fixing a
point $x$ of the orbit bundle $\Sigma_{(F)}$. This amounts to a
special treatment of possible symmetry axes or centers,
respectively.

By restricting ourselves to one fixed orbit bundle we  fix an
isotropy subgroup $F\leq S$ and we require that the action of $S$
on $\Sigma$ is such that the orbits are given by $S(x)\cong S/F$
for all $x\in\Sigma$. This will be the case if $S$ is compact.
Moreover, we have to assume that the coset space $S/F$ is
reductive \cite{KobNom}, i.e. ${\cal L}S$ can be decomposed as a
direct sum ${\cal L}S={\cal L}F\oplus{\cal
  L}F_{\perp}$ with $\Ad_F({\cal L}F_{\perp})\subset{\cal
  L}F_{\perp}$. If $S$ is semisimple, ${\cal L}F_{\perp}$ is the
orthogonal complement of ${\cal L}F$ with respect to the
Cartan-Killing metric on ${\cal L}S$. The base manifold can then
be decomposed as $\Sigma\cong\Sigma/S\times S/F$ where
$\Sigma/S\cong B\subset\Sigma$ is the base manifold of the orbit
bundle and it can be realized as a submanifold $B$ of $\Sigma$ via
a section in this bundle.

Given a point $x\in\Sigma$, the action of the isotropy subgroup
$F$ yields a map $F\colon\pi^{-1}(x)\to\pi^{-1}(x)$ of the fibre
over $x$ commuting with the right action of the bundle.  To each
point $p\in\pi^{-1}(x)$ we can assign a group homomorphism
$\lambda_p\colon F\to G$ defined by $f(p)=:p\cdot\lambda_p(f)$ for
all $f\in F$. For a different point $p^{\prime}=p\cdot g$ in the
same fibre we get, using commutativity of the action of
$S<\Aut(P)$ with right multiplication of $G$ on $P$, the
conjugated homomorphism
$\lambda_{p^{\prime}}=\Ad_{g^{-1}}\circ\lambda_p$. This
construction yields a map $\lambda\colon P\times F\to
G,(p,f)\mapsto\lambda_p(f)$ obeying the relation $\lambda_{p\cdot
g}=\Ad_{g^{-1}}\circ\lambda_p$.

Given a fixed homomorphism $\lambda\colon F\to G$, we can build
the principal fibre subbundle
\begin{equation}
 Q_{\lambda}(B,Z_{\lambda},\pi_Q):=\{p\in P_{|B}:\lambda_p=\lambda\}
\end{equation}
over the base manifold $B$ the structure group of which is the
centralizer $Z_{\lambda}:=Z_G(\lambda(F))$ of $\lambda(F)$ in $G$.
$P_{|B}$ is the restricted fibre bundle over $B$. A conjugated
homomorphism $\lambda^{\prime}=\Ad_{g^{-1}}\circ\lambda$ will lead
to an isomorphic fibre bundle.

The structure elements $[\lambda]$ and $Q$ classify symmetric
principal fibre bundles according to the following theorem
\cite{Brodbeck}:

\begin{theo}\label{bundle}
  A $S$-symmetric principal fibre bundle $P(\Sigma,G,\pi)$ with the
  iso\-tropy subgroup $F\leq S$ of the action of $S$ on $\Sigma$ is
  uniquely characterized by a conjugacy class $[\lambda]$ of
  homomorphisms $\lambda\colon F\to G$ together with a {\em reduced bundle}
  $Q(\Sigma/S,Z_G(\lambda(F)),\pi_Q)$.
\end{theo}

Given two groups $F$ and $G$ we can make use of the relation
\cite{BroeckerDieck}
\begin{equation}\label{Hom}
 \Hom(F,G)/\Ad\cong\Hom(F,T(G))/W(G)
\end{equation}
in order to determine all conjugacy classes of homomorphisms
$\lambda\colon F\to G$. Here $T(G)$ is a standard maximal torus
and $W(G)$ the Weyl group of $G$.

Now let $\omega$ be a $S$-invariant connection on the bundle $P$
classified by $([\lambda],Q)$. The connection $\omega$ induces a
connection $\tilde{\omega}$ on the reduced bundle $Q$. Because of
the $S$-invariance of $\omega$ the reduced connection
$\tilde{\omega}$ is a one-form on $Q$ with values in the Lie
algebra of the reduced structure group. Furthermore, by using
$\omega$ we can construct the linear map $\Lambda_p\colon\Lie
S\to\Lie G,X\mapsto\omega_p(\tilde{X})$ for any $p\in P$. Here
$\tilde{X}$ is the vector field on $P$ given by $\tilde{X}(f):=
d(\exp(tX)^{\star}f)/dt|_{t=0}$ for any $X\in\Lie S$ and $f\in
C^1(P,\R)$.  For $X\in\Lie F$ the vector field $\tilde{X}$ is a
vertical vector field, and we have $\Lambda_p(X)=\md\lambda_p(X)$
where $\md\lambda\colon\Lie F\to\Lie G$ is the derivative of the
homomorphism defined above. This component of $\Lambda$ is
therefore already given by the classifying structure of the
principal fibre bundle. Using a suitable gauge, $\lambda$ can be
held constant along $B$. The remaining components
$\Lambda_p|_{{\cal L}F_{\perp}}$ yield information about the
invariant connection $\omega$. They are subject to the condition
\begin{equation}\label{Higgs}
 \Lambda_p(\Ad_f(X))=\Ad_{\lambda_p(f)}(\Lambda_p(X))\quad
 \mbox{ for }f\in F,X\in{\cal L}S
\end{equation}
which follows from the transformation of $\omega$ under the
adjoint representation and which provides a set of equations which
determine the Higgs field.

Keeping only the information characterizing $\omega$ we have,
besides $\tilde{\omega}$, the Higgs field $\phi\colon Q\to\Lie
G\otimes\Lie F_{\perp}^{\star}$ determined by $\Lambda_p|_{{\cal
L}F_{\perp}}$.  The reduced connection and the Higgs field suffice
to characterize an invariant connection. This is the assertion of
the following theorem \cite{Brodbeck}:

\begin{theo}[Generalized Wang theorem]\label{connect}
  Let $P(\Sigma,G)$ be a $S$-sym\-me\-tric principal fibre bundle
  classified by $([\lambda],Q)$ according to Theorem~\ref{bundle} and
  let $\omega$ be a $S$-invariant connection on $P$.

  Then the connection $\omega$ is uniquely
  classified by the {\em reduced connection}
  $\tilde{\omega}$ on $Q$ and the Higgs field $\phi\colon Q\times\Lie
  F_{\perp}\to\Lie G$ obeying Eq.\ (\ref{Higgs}).
\end{theo}

In general, $\phi$ will transform under some representation of the
reduced structure group $Z_{\lambda}$: The Higgs field lies in the
subspace of ${\cal L}G$ determined by Eq.\ (\ref{Higgs}).  It
forms a representation space of all group elements of $G$ (which
act on $\Lambda$) whose action preserves the Higgs subspace. These
are precisely elements of the reduced group by definition.

The connection $\omega$ can be reconstructed from its classifying
structure as follows: According to the decomposition $\Sigma\cong
B\times S/F$ we have $\omega=\tilde{\omega}+\omega_{S/F}$, where
$\omega_{S/F}$ is given by $\Lambda\circ\iota^{\star}\theta_{MC}$
in a gauge depending on the (local) embedding $\iota\colon
S/F\hookrightarrow S$. Here $\theta_{MC}$ is the Maurer-Cartan
form on $S$.  For example, in the generic case (not in a symmetry
center) of spherical symmetry we have $S=SU(2)$,
$F=U(1)=\exp\langle\tau_3\rangle$ ($\langle\cdot\rangle$ denotes
the linear span), and the connection form can be gauged to be
\begin{equation}\label{ASF}
 A_{S/F}=(\Lambda(\tau_2)\sin\vartheta+\Lambda(\tau_3)\cos\vartheta)
  \md\varphi+\Lambda(\tau_1)\md\vartheta.
\end{equation}
Here $(\vartheta,\varphi)$ are coordinates on $S/F\cong S^2$. The
$\tau_j$ build a basis of $\Lie S$ and are given by
$\tau_j:=-\frac{i}{2}\sigma_j$, $\sigma_j$ being the Pauli
matrices. $\Lambda(\tau_3)$ is given by $\md\lambda$, whereas
$\Lambda(\tau_{1,2})$ are the Higgs field components.

Eq.\ (\ref{ASF}) contains as special cases the invariant
connections found in Ref.\ \cite{Cordero}.  These are gauge
equivalent by gauge transformations depending on the angular
coordinates $(\vartheta,\varphi)$, i.~e., they correspond to
homomorphisms $\lambda$ which are not constant on the orbits of
the symmetry group.

\section{Symmetric States}\label{s:SymmState}

Before describing the rather abstract construction of symmetric
spin network states we will present the general idea in a first
subsection. The following subsections deal with the construction
of symmetric states as generalized states of the unreduced theory
and proofs of some of their properties.

\subsection{Principal Idea}

The principal idea of our construction \cite{MBDiplom} described
in this section is to make use of the reconstruction of an
invariant connection from its classifying structure, namely by
means of the pull back of a function on the space of connections
on $\Sigma$ to a function on the space of connections plus Higgs
fields on the reduced manifold $B$ which in the context of
analytic spin networks will be assumed to be an analytic
submanifold of $\Sigma$.

But some complications arise because of the classical partial
gauge fixing by selecting a special homomorphism
$\lambda\in[\lambda]$. The reduced gauge group and the space of
Higgs fields depend on this selection. Moreover, the Higgs field
does in general not transform under the adjoint representation of
the reduced structure group, which would be helpful in spin
network quantization. In contrast, before imposing the constraint
(\ref{Higgs}) it transforms in general under the adjoint
representation of the {\em unreduced} structure group. Such a
Higgs field can easily be implemented in the spin network context
using the rules recalled in Subsection~\ref{s:spinnet}.

The interrelation of partial gauge fixings and  reductions of the
gauge group makes it possible to eliminate partial gauge fixings
by using the full gauge group on the reduced manifold. This is the
essence of definition~\ref{AUG} below.

\subsection{Construction}
\label{s:construct}

Let us now define the notion of symmetric states. The Hilbert
space under consideration is the auxiliary Hilbert space ${\cal
H}_{\Sigma}:=L_2(\AbS,\md\mu_{AL})$ on which the constraints have
to be solved.  Because of the singular character of symmetric
states mentioned in the introduction we will have to use the
rigged Hilbert space $\Phi_{\Sigma}\subset{\cal
H}_{\Sigma}\subset\Phi_{\Sigma}^{\prime}$, where $\Phi_{\Sigma}$
denotes the space of cylindrical functions on the space $\AbS$ of
connections over $\Sigma$ and $\Phi_{\Sigma}^{\prime}$ its
topological dual.

\begin{defi}[Symmetric States]\label{SymmState}
  Let $P$ be a $S$-symmetric principal fibre bundle, classified by
  $([\lambda],Q)$.

  A {\em $[\lambda]$-symmetric state} is a distribution
  $\psi\in\Phip_{\Sigma}$ on $\Phi_{\Sigma}$ whose support contains
  only connections that are invariant under the $S$-action on $P$
  classified by $[\lambda]$.
\end{defi}

Although Definition~\ref{SymmState} catches the intuitive notion
of a symmetric state, it is not well suited for a calculus. We
have to develop some tools in analogy to the spin network
calculus. This will be done in the remaining part of this section
by combining techniques collected in the last section. Application
of these techniques will lead to several spaces of connections
which are defined in

\begin{defi}
  Let $P(\Sigma,G)$ be a principal fibre bundle acted on by a symmetry
  group $S$ according to the classification $([\lambda],Q)$, where $Q$
  is the reduced bundle over the manifold $B\subset\Sigma$.

  $\AbS$ and $\GbS$ are the space of generalized connections on $P$
  and the extended local gauge group, respectively. $\AbR\times\UbR$
  is the space of generalized $G$-connections and Higgs fields in the
  adjoint representation of $G$ over $B$. $\GbR$ is the extended local
  gauge group of generalized $G$-gauge transformations over $B$.

  For any $\lambda^{\prime}\in[\lambda]$,
  $(\Ab\times\Ub)^{\lambda^{\prime}}$ is defined to be the subset of
  $\AbR\times\UbR$ subject to the following constraints: The
  generalized connections take values in the structure group
  $Z_{\lambda^{\prime}}$ of the reduced bundle $Q$, and the
  generalized Higgs fields take values in the submanifold of $G$
  obtained by exponentiating the linear solution space of
  Eq.\ (\ref{Higgs}). Here we have to use a separate Higgs field
  component for every element of a basis of $\Lie F_{\perp}$.
\end{defi}

\begin{rem}
  The space $\AbR\times\UbR$ is independent of the reduction of the
  structure group and the constraints (\ref{Higgs}) on the Higgs
  field. These affect only the definition of
  $(\Ab\times\Ub)^{\lambda^{\prime}}$ which depends explicitly on the
  homomorphism $\lambda^{\prime}$, not only on its conjugacy class.
  Therefore,  for any $\lambda^{\prime}\in[\lambda]$ we have a separate
  space of connections and Higgs fields because already the reduced
  structure group may depend on $\lambda^{\prime}$. We can eliminate this
  redundancy by factoring out the gauge group, but this has to be done
  with care due to the classical reduction of the gauge group.
\end{rem}

In order to achieve our goal, we will make use of the classifying
structure $(\tilde{\omega},\phi)$ of a $[\lambda]$-invariant
connection $\omega$.  Below Theorem~\ref{connect}, we described
the reconstruction of $\omega$ from its classifying structure.
This reconstruction defines a continuous map
$$
  r_{\lambda^{\prime}}^{(\iota)}\colon({\cal A}\times{\cal U})^{\lp}\to\AbS
$$
which can be continued uniquely to a continuous map
$$
  r_{\lambda^{\prime}}^{(\iota)}\colon(\Ab\times\Ub)^{\lp}\to\AbS.
$$
As the notation indicates, this map depends not only on the
homomorphism $\lp\in[\lambda]$, but also on the embedding
$\iota\colon S/F\hookrightarrow S$.

Because a different $\iota$ would reconstruct a gauge equivalent
connection form, the dependence on $\iota$ can be eliminated by
factoring out the gauge group on $P$. This leads us to the family
of maps
\begin{equation}\label{rlp}
  r_{\lp}\colon (\Ab\times\Ub)^{\lp}\to\AbS/\GbS.
\end{equation}
The dependence on $\lp$ (as opposed to $[\lambda]$) is pure gauge:
$\lp$ can be changed arbitrarily in its conjugacy class
$[\lambda]$ by applying a global transformation with a $g\in G$,
$g\not\in Z_{\lambda}$.  This shows that the domains of definition
of all the maps $r_{\lp}$ are in fact different, but that they are
related by gauge transformations. This observation motivates the
following

\begin{defi}\label{AUG}
  Let $[\lambda]$ be a conjugacy class of homomorphisms.

  Then $(\Ab\times\Ub/\Gb)^{[\lambda]}$ is the subset of
  $\AbR\times\UbR/\GbR$ consisting of all $\GbR$-gauge equivalence
  classes containing a representative which lies in some
  $(\Ab\times\Ub)^{\lp}$, $\lp\in[\lambda]$.
\end{defi}

\begin{rem}
  Because we allow here for any local gauge transformation, we relax
  the condition that $\lp$ be constant on $B$, which is imposed in the
  classical symmetry reduction procedure.
\end{rem}

$\Gb$-Equivariance of $r_{\lp}$, which means that for any
$\GbR$-gauge transformation $g_B$ there is a $\GbS$-gauge
transformation $g_{\Sigma}$ with $r_{\lp}\circ g_B=g_{\Sigma}\circ
r_{\lp}$, now allows to factor out gauge transformations in the
domains of definition of the maps $r_{\lp}$. Thereby we obtain a
further map
\begin{equation}\label{rclp}
  r_{[\lambda]}\colon(\Ab\times\Ub/\Gb)^{[\lambda]}\to\AbS/\GbS
\end{equation}
which  depends only on the conjugacy class $[\lambda]$.

We then have the

\begin{lemma}\label{imager}
  The subset of generalized gauge invariant, $[\lambda]$-invariant
  connections in $\AbS/\GbS$ is given by $\Ima(r_{[\lambda]})$, the
  image of the map $r_{[\lambda]}$.
\end{lemma}

\begin{proof}
 This is clear from the construction of $r_{[\lambda]}$.
\end{proof}

Recall that our goal is to develop a calculus on the manifold of
$[\lambda]$-invariant connections.  Given a continuous function on
this manifold, we can pull it back via $r_{[\lambda]}$ and so
obtain a continuous function on $(\Ab\times\Ub/\Gb)^{[\lambda]}$.
If this function can be continued to a function on
$\AbR\times\UbR$, we will have the desired calculus on that space
at our disposal. An extension can indeed be achieved by expanding
the pulled back function in the spin network basis of
$C(\AbR\times\UbR)$, the space of continuous functions on the
space of connections and Higgs fields over $B$.

In order to achieve uniqueness of this expansion we have to
truncate the spin network basis ${\cal B}$ of $C(\AbR\times\UbR)$
if necessary in such a way that
\begin{equation}
  \hat{{\cal B}}:=\{T|_{(\Ab\times\Ub)^{[\lambda]}}\}_{T\in{\cal B}}
\end{equation}
is a set of independent functions. E.g. we have to use only spin
network states with trivial Higgs vertices if Eq.\ (\ref{Higgs})
does not allow  any nonvanishing Higgs field.

This extension procedure following the pull back with
$r_{[\lambda]}$ finally yields the map
\begin{equation}\label{rstar}
  \rstar\colon C(\AbS/\GbS)\to C(\AbR\times\UbR/\GbR)
\end{equation}
which will provide the key element in our investigation of
symmetric states.

However, a function pulled back in such a way is quite singular on
the space $\AbR\times\UbR/\GbR$. Even if we constrain the domain
of definition of $\rstar$ to $\Phi_{\Sigma}$, the space of
cylindrical functions, the pull back may in general not lead to a
cylindrical or $AL$-integrable function on $\AbR\times\UbR/\GbR$:
The holonomy to a generic edge depends on all components of an
unreduced connection in all its points, which leads to a
continuous distribution of Higgs vertices.  We, therefore, have
again to use a rigged Hilbert space, this time
\begin{equation}\label{RiggedB}
  \Phi_B\subset{\cal H}_B\subset\Phi_B^{\prime}.
\end{equation}
Here, $\Phi_B$ is the space of cylindrical functions on
$\AbR\times\UbR/\GbR$, $\Phi_B^{\prime}$ its topological dual, and
${\cal H}_B:=L_2(\AbR\times\UbR/\GbR,\md\mu_{AL})$ (again modulo
relations which solve the Higgs constraint (\ref{Higgs}), and
which will be dealt with in more detail elsewhere; examples can be
found in the last section).

 The restriction of $\rstar$ to
$\Phi_{\Sigma}$ can now be interpreted as an antilinear map
$$
  \rho_{[\lambda]}\colon\Phi_{\Sigma}\to\Phi_B^{\prime},
$$
reminiscent of a group averaging map: The pull back of a
cylindrical function $f\in\Phi_{\Sigma}$ is interpreted as a
distribution on $\Phi_B$ according to ($g\in\Phi_B$)
\begin{equation}
  \rho_{[\lambda]}(f)(g):=\int_{\AbR\times\UbR/\GbR}\md\mu_{AL}
  \:\overline{\rstar f}\:g.
\end{equation}

In a similar way, we can interpret a cylindrical function
$g\in\Phi_B$ as a distribution on $\Phi_{\Sigma}$ according to
\begin{equation}\label{sigma}
  \sigma_{[\lambda]}(g)(f):=\int_{\AbR\times\UbR/\GbR}\md\mu_{AL}
   \:\overline{g}\:\rstar f
\end{equation}
where $\sigma_{[\lambda]}\colon\Phi_B\to\Phi_{\Sigma}^{\prime}$ is
the antilinear map given by this interpretation.

The situation can be summarized in the diagram
\setlength{\unitlength}{1.5mm}
\begin{center}\begin{picture}(30,30)
 \put(5,25){\makebox(0,0)[t]{$\Phi_B^{\prime}$}}
 \put(5,20){\makebox(0,0)[t]{$\bigcup$}}
 \put(5,15){\makebox(0,0)[t]{${\cal H}_B$}}
 \put(5,10){\makebox(0,0)[t]{$\bigcup$}}
 \put(5,5){\makebox(0,0)[t]{$\Phi_B$}}
 \put(25,25){\makebox(0,0)[t]{$\Phi_{\Sigma}^{\prime}$}}
 \put(25,20){\makebox(0,0)[t]{$\bigcup$}}
 \put(25,15){\makebox(0,0)[t]{${\cal H}_{\Sigma}$}}
 \put(25,10){\makebox(0,0)[t]{$\bigcup$}}
 \put(25,5){\makebox(0,0)[t]{$\Phi_{\Sigma}$}}
 \put(7,6){\vector(1,1){16}}
 \put(23,6){\vector(-1,1){16}}
 \put(10,7.5){$\sigma_{[\lambda]}$}
 \put(10,20){$\rho_{[\lambda]}$}
\end{picture}\end{center}
with the duality relation
\begin{equation}
 \sigma_{[\lambda]}(\overline{g})(f)=\rho_{[\lambda]}(\overline{f})(g)
 \quad\mbox{ for }f\in\Phi_{\Sigma},g\in\Phi_B
\end{equation}
between the maps $\sigma_{[\lambda]}$, $\rho_{[\lambda]}$
connecting the two Gel'fand triples.

In general (see the remarks preceding Eq.\ (\ref{RiggedB}))
$\rho_{[\lambda]}(\Phi_{\Sigma})$ is not contained in $\Phi_B$,
and we cannot compose $\rho_{[\lambda]}$ and $\sigma_{[\lambda]}$
to obtain a map from $\Phi_{\Sigma}$ to $\Phi_{\Sigma}^{\prime}$.
This is the main difference to a group averaging map, which is
aimed to solve a gauge constraint. Only in very special situations
can the symmetry reduction be formulated analogous to a group
averaging (Subsection~\ref{s:SphSymmElec} and \cite{BFElec}). But
in general we have the two maps $\rho_{[\lambda]}$, which
restricts an unsymmetric state to its symmetric part, and
$\sigma_{[\lambda]}$, which identifies symmetric states with spin
network states over $B$ thereby equipping the space of symmetric
states with a calculus.

\subsection{Properties of the Symmetric States}

The goal of this subsection is to prove that the construction of
the previous subsection yields all symmetric states. In order to
achieve this we need some preparations.

\begin{lemma}\label{centralizer}
  Let $G$ be a compact topological group which is Hausdorff and $H$ be
  a subgroup of $G$. The centralizer $Z_G(H):=\{g\in G:gh=hg\mbox{ \rm
    for all }h\in H\}$ is a compact subgroup of $G$.
\end{lemma}

\begin{proof}
  It is well known that $Z_G(H)$ is a subgroup \cite{BroeckerDieck}. It
  can be written as
 $$  Z_G(H)=\bigcap_{h\in H}G_h  $$
  where $G_h$ is the isotropy subgroup of $h\in H$ under the adjoint
  action $G\times G\to G$, $(g,h)\mapsto ghg^{-1}$ of $G$ on
  itself. Because the intersection of an arbitrary set of closed sets
  is closed, it suffices to prove that all isotropy subgroups $G_h$
  are closed.

  The action of $G$ on itself leads, restricted to a fixed element
  $h\in G$, to the continuous map $c_h\colon G\to G,g\mapsto
  ghg^{-1}$. $G_h=c_h^{-1}(h)$ is the preimage of a closed set
  (because $G$ is Hausdorff) under a continuous map, and hence
  closed. Now, $Z_G(H)$ is a closed subset of a compact group and
  therefore compact.
\end{proof}

\begin{lemma}\label{invconn}
  Let $P$ be a $S$-symmetric principal fibre bundle and $[\lambda]$ be
  a conjugacy class of homomorphisms classifying $P$ together with the
  reduced bundle $Q$.

  The set of generalized gauge invariant $[\lambda]$-invariant
  connections on $P$ is closed in $\AbS/\GbS$.
\end{lemma}

\begin{proof}
  According to Lemma~\ref{imager} we have to show that the image of
  $r_{\lambda}$ --~which is identical to the image of
  $r_{[\lambda]}$~-- is closed.

  We will start by showing that the domain of definition of
  $r_{\lambda}$, i.~e.\ $(\Ab\times\Ub)^{\lambda}$, is compact. The
  elements of this space take values in a compact set, because we know
  in the first place from Lemma~\ref{centralizer} that $Z_{\lambda}$
  is compact. The generalized Higgs fields take values in a compact
  manifold, too. Such a Higgs field has several components given by the
  linear map $\phi\colon\Lie F_{\perp}\to\Lie G$ subject to the linear
  Eq.\ (\ref{Higgs}). Therefore, the values of $\phi$ lie in a
  linear subspace of $\Lie G$, and exponentiating yields generalized
  Higgs fields taking values in a compact submanifold of $G$.

  $(\Ab\times\Ub)^{\lambda}$ can be constructed as a projective limit
  along the lines described in Subsection~\ref{s:spinnet}. The
  projective family consists of compact spaces, because the maps
  involved assign elements of compact spaces to a finite number of
  edges and vertices of graphs. Therefore, the projective limit is
  compact in its induced Tychonov topology which is equivalent to the
  induced topology as a subset of $\AbR\times\UbR$.

  Now, as stated in Subsection~\ref{s:spinnet}, the Tychonov topology
  is equivalent to the Gel'fand topology. But $r_{\lambda}$ is
  continuous in the Gel'fand topology (being by construction a
  continuation of a continuous function on the dense subspace $({\cal
  A}\times{\cal U})^{\lambda}$), and so the image of $r_{\lambda}$ is
  compact. As a compact subset of a Hausdorff space it is closed.
\end{proof}

\begin{theo}\label{sigl}
  Let $[\lambda]$ be a conjugacy class of homomorphisms.

  The image of $\sigma_{[\lambda]}$ contains only $[\lambda]$-symmetric states.
\end{theo}

\begin{proof}
  According to Definition~\ref{SymmState} we have to prove that the
  support of a distribution in the image of $\sigma_{[\lambda]}$
  contains only $[\lambda]$-invariant connections. This amounts to
  showing that for any non-invariant generalized connection
  $\overline{A}\in(\AbS/\GbS)\backslash\Ima(r_{[\lambda]})$ there is a
  neighborhood $U$ of $\overline{A}$ so that the restriction of any
  distribution $\psi\in\Ima(\sigma_{[\lambda]})$ to $U$ is the zero
  distribution.

  Because of Lemma~\ref{invconn} $\overline{A}$ has a neighborhood
  which is entirely contained in $(\AbS/\GbS)\backslash\Ima(r_{[
  \lambda]})$. If
  we restrict $\psi$ to this neighborhood, it will be the zero
  distribution due to its very definition in Eq.\ (\ref{sigma}):
  The pull back with $r_{[\lambda]}$ of any function supported on the
  neighborhood of $\overline{A}$ will be zero.
\end{proof}

This theorem provides us with a rich class of
$[\lambda]$-symmetric states. Even better, we have a calculus on
this class of states, because they are identified by Eq.\
(\ref{sigma}) with the space $\Phi_B$ of cylindrical functions.
Elements of $\Phi_B^{\prime}$ can be regarded as generalized
symmetric states.

The following lemma states that we have found enough symmetric
states.

\begin{lemma}\label{separate}
  Let $[\lambda]$ be a conjugacy class of homomorphisms.

  There is no cylindrical function $f\in\Phi_{\Sigma}$ that is
  non-trivial on $\Ima(r_{[\lambda]})$ and annihilated by all
  distributions in $\sigl(\Phi_B)$. In particular, the space
  $\sigl(\Phi_B)$ separates the elements of $\Phi_{\Sigma}$ that
  differ when restricted to $[\lambda]$-invariant connections.
\end{lemma}

\begin{proof}
  Let $\psi\in\Phi_{\Sigma}^{\prime}$ be a symmetric state. If
  $f,g\in\Phi_{\Sigma}$ are two cylindrical functions that are
  identical when restricted to $[\lambda]$-invariant connections, then
  $\psi(f)=\psi(g)$. Let us, therefore, introduce the following
  equivalence relation which respects the algebraic and topological
  structure of $\Phi_{\Sigma}$.  Two cylindrical functions $f\sim g$
  are equivalent if and only if $\rstar f=\rstar g$.  Symmetric states
  can be seen as functions on the space of equivalence classes, and we
  need to show that, if $[f]\not=[g]$, there is a distribution
  $\psi\in\sigl(\Phi_B)$ with $\psi(f)\not=\psi(g)$.

  Let us look now at the function $\rstar f\in C(\AbR\times\UbR/\GbR)$
  where $[f]$ is not trivial. As noted earlier, it may not be
  cylindrical, not even integrable. But if it is cylindrical, it will
  correspond to a symmetric state obeying $\sigl(\rstar
  f)(f)\not=0$. If it is not, we can approximate it by a sequence of
  cylindrical functions which is obtained by projecting it onto
  cylindrical subspaces, cylindrical with respect to graphs of an
  increasing net constructed as follows: If $\rstar f$ will lie in
  $\Phi_B^{\prime}$, but not in $\Phi_B$, then it will be a countably
  infinite sum of terms, each being cylindrical with respect to a
  finite graph, the union of all these graphs being an infinite
  graph. The projections will be obtained by truncating to a finite
  number of these graphs, their number tending to infinity in the
  sequence mentioned.  (This is reminiscent of the well known
  approximating sequences of the $\delta$-distribution or, more
  generally, of an approximate identity in an algebra without
  identity.)  All the
  functions $f_i$ in the sequence will fulfill $\sigl(f_i)(f)\not=0$
  proving our assertion.

  Separation now follows from linearity, but can also be proved
  directly: We pick a representative for each
  equivalence class of cylindrical functions in
  $\Phi_{\Sigma}$, and for each representative $f$ the cylindrical
  function $\rstar f$, if it lies already in $\Phi_B$, or else an
  appropriate element of the sequence approximating $\rstar f$.  The
  term appropriate means, that any linear relation between the chosen
  functions reflects a linear relation between equivalence
  classes. An appropriate selection can always be done, because the
  sequences approximate the functions $\rstar f$ which are different
  for different classes. Thereby, we obtain a class of
  distributions separating the equivalence classes.
\end{proof}

The results of this lemma can be interpreted as
(over-)completeness of the set $\sigma_{[\lambda]}(\Phi_B)$ of
generalized functions on
$\{f|_{\Ima(r_{[\lambda]})}:f\in\Phi_{\Sigma}\}$ in the sense that
$\sigma_{[\lambda]}(g)(f)=0$ for all $g\in\Phi_B$ implies
$f|_{\Ima(r_{[\lambda]})}=0$. Together with Theorem~\ref{sigl}
this can be summarized in

\begin{theo}[Quantum Symmetry Reduction]\label{ident}

  Let $P$ be a $S$-symmet\-ric principal fibre bundle classified by
  $([\lambda],Q)$, $Q$ being the reduced bundle over $B$.

  The space of $[\lambda]$-symmetric states on $\AbS/\GbS$ can, by means of
the mapping $\sigl$, be identified with the space $\Phi_B$ of
cylindrical
  functions on $\AbR\times\UbR/\GbR$.
\end{theo}

\begin{proof}
  Let $\Phi_{symm}:=\Phi_{\Sigma}/\sim$ be the space of functions on
  the space $\Ima(r_{[\lambda]})$ of generalized gauge invariant
  and
  $[\lambda]$-invariant connections, and $\Phi_{symm}'$ its
  topological dual. Due to the definition in Lemma~\ref{separate} of
  $\sim$, $\Phi_{symm}'$ can be identified with the space of
  $[\lambda]$-symmetric states as defined in
  Definition~\ref{SymmState}.

  According to Lemma~\ref{separate}, $\sigma_{[\lambda]}(\Phi_B)$ is in
  separating duality with $\Phi_{symm}$. This implies that
  $\sigma_{[\lambda]}(\Phi_B)$ is dense in $\Phi_{symm}'$ in the weak
  topology \cite[Chapter II, \S 6.2, Corollary 4]{Bourbaki:TopVec},
  and, therefore, in the space of $[\lambda]$-symmetric states. (Note,
  however, that the topology on $\Phi_B$ induced from the weak
  topology by the continuous map $\sigma_{[\lambda]}$ is coarser than
  the topology in which $\Phi_B$ is completed to ${\cal H}_B$.)

  In conclusion, the map $\sigma_{[\lambda]}$ is injective (by construction of
  $\Phi_B$ -- where the Higgs constraint (\ref{Higgs}) is assumed to
  be solved -- and by the construction of $\sigma_{[\lambda]}$),
   and has a dense image in the
  space of $[\lambda]$-symmetric states.
\end{proof}

This theorem allows us to trade the space of symmetric states
according to Definition~\ref{SymmState}, which is not well suited
for establishing a calculus, for the space $\Phi_B$. This space
and the given calculus thereon are only a bit more difficult to
deal with than the space $\Phi_{\Sigma}$ of all states of the full
theory, because we may have to use spin networks with Higgs field
vertices.

The Gau\ss\ constraint is solved by using gauge invariant
functions, i.e.\ cylindrical functions on $\AbR\times\UbR/\GbR$.
The diffeomorphism constraint can be solved by group averaging.
After imposing $S$-symmetry there are only those diffeomorphisms
to average that respect this symmetry. These are precisely the
diffeomorphisms of $B$, and the constraint can be solved by
averaging over the diffeomorphism group of $B$ acting on $\Phi_B$.

We conclude this section with some remarks:

\begin{itemize}
 \item We describe symmetric states by spin networks of the group $G$,
  not those of $Z_{\lambda}$, as might have been expected from the classical
  reduction. On the one hand, we are forced to do that because the
  Higgs field transforms in general according to the adjoint
  representation of $G$, not $Z_{\lambda}$. On the other hand, the
  reduced structure group $Z_{\lambda}$ has its origin in a partial
  gauge fixing imposed by fixing $\lambda$. Our quantum symmetry
  reduced theory does not make use of such a partial gauge fixing.
  Instead we will
  have to enforce gauge invariance under the full group $G$.

 \item In quantizing classical expressions we have nevertheless to
  start from those which might be invariant only under the
  reduced gauge group. The first step of quantization will yield an
  operator acting on functions on $(\Ab\times\Ub)^{\lambda}$. It has
  then to be extended by gauge covariance to an operator on $\Phi_B$. This
  is parallel to our construction of the map $r_{[\lambda]}$ in the
  equations~(\ref{rlp}), (\ref{rclp}) and (\ref{rstar}).  An example
  of the quantization of such an operator  will be given in
  Subsection~\ref{Gauss}.

 \item In order to be actually able to quantize classical expressions
  in the Hamiltonian formalism used so far, we will have to claim
  validity of the symmetric criticality principle in the theory under
  consideration. This means that critical points of the reduced
  action should correspond to critical points of the unreduced one. Validity of
  this principle cannot be taken for granted. It will be valid, if
  the symmetry group $S$ is compact, or if it is abelian (more
  generally, unimodular) and acts freely \cite{midisup}. These two
  cases cover our examples given below.

 \item Until now we have confined ourselves to a fixed conjugacy class
  $[\lambda]$. In general there will be a family of such conjugacy
  classes. If it is not possible to select one by means of physical
  considerations, we will have to treat them on an equal footing. We
  will use the same space $\Phi_B$ for every class, but they will be
  differently represented as distributions on $\Phi_{\Sigma}$ by means
  of $\sigl$. The different conjugacy classes give rise to different
  sectors, which are superselected as seen from the symmetric
  theory. They correspond to different subspaces of invariant
  connections embedded in $\AbS/\GbS$ via $r_{[\lambda]}$.  We have to
  find a physical interpretation for the different conjugacy classes,
  the most natural being that of a topological charge. This is purely
  classical, because the different conjugacy classes arise already in
  the classical reduction. We will see in the next section that this case
  indeed occurs, namely for
   spherically symmetric electromagnetism where $[\lambda]$
  gives the magnetic charge.

 \item Classically a symmetric principal fibre bundle is classified
  by a conjugacy class of homomorphisms and a reduced bundle, see
  Theorem~\ref{bundle}. In the quantum theory, however, the conjugacy
  class of the homomorphisms will suffice, because the space $\AbR$ of
  generalized connections contains connections on all the bundles over
  $B$, as proven in \cite{AL:HolAlg}.

 \item The emergence of the Higgs field can be understood as
  a reflection of the information which is contained in those edges
  associated with a cylindrical state which
  are located  entirely in the orbits of $S$.  E.g. in a spherically
  symmetric theory they represent edges in a $S^2$ orbit. This
  justifies even more the name ``point holonomies''. From this picture
  one can see that in general the function $\rstar f$ will involve an
  infinite number of Higgs vertices, where $f\in\Phi_{\Sigma}$ is a
  cylindrical function. If the underlying graph contains an edge that
  lies neither entirely in an orbit nor entirely in the manifold $B$,
  then $f$ will depend on all components of a connection in points
  along the edge. In the symmetry reduced theory some of the
  components are given by the Higgs field, so the full dependence of
  $f$ on a connection requires a continuous set of Higgs vertices in
  the union of all graphs underlying $\rstar f$.

 \item The procedure of the present section can be reversed, using a
  Kaluza-Klein construction of a Higgs field as components of a
  connection in a higher-dimensional, symmetric manifold.  Suppose we
  want to quantize a diffeomorphism invariant theory of a connection
  and a Higgs field, for which there is a higher-dimensional
  Kaluza-Klein theory containing only connection fields.
   Then we can visualize the
  Higgs vertices arising in the quantization of the lower-dimensional
  theory as remnants of loops lying in the compactified extra
  dimensions. By ``blowing up'' a lower-dimensional spin network with
  Higgs vertices in this way in order to obtain a higher-dimensional
  ordinary spin
  network, we get a quantization of the Kaluza-Klein theory. This
  blowing up preserves, like our symmetry reduction, gauge invariance,
  because the Higgs vertices transform under the adjoint
  representation of the structure group as do loops based on the
  vertices.
\end{itemize}

\section{Examples}\label{s:Examples}

In this section we will give examples in order to illustrate some
of the ideas of the last section  and to test them.

\subsection{\boldmath $2+1$ Dimensional Gravity}

$2+1$ dimensional gravity can be obtained in terms of a symmetry
reduction of $3+1$ dimensional gravity by imposing the existence
of one spacelike Killing vector field of constant norm. The extra
condition on the norm of the Killing field is necessary in order
to eliminate a scalar field which is related to its norm.
Otherwise one would obtain $2+1$ dimensional gravity coupled to a
massless scalar field.

We therefore have the abelian symmetry group $S=\R$ (or
$S=SO(2)$), and we assume the space $\Sigma$ to have the topology
$\Sigma=B\times\R$ (or $\Sigma=B\times S^1$) where $B$ is a
two-dimensional manifold. The group $S$ acts freely on the second
component of $\Sigma$  which means that $F=\{0\}$. Hence, there
can only be one homomorphism $\lambda\colon F\to G,\, 0\mapsto 1$.
Here $G=SU(2)$ is the gauge group of gravity, formulated as a
gauge theory using real Ashtekar variables. In this case the
reduced group $Z_{\lambda}=G$ is identical to the full structure
group. Because $F$ is trivial, Eq.\ (\ref{Higgs}) is trivially
true and we have a one-component Higgs field $\phi$ taking values
in $\Lie G$. If we use the Maurer-Cartan form $\theta=\md z$,
where $z$ coordinatizes $\R$, and the embedding $\iota=\id\colon
\R\to\R$, the reconstruction is
\begin{equation}\label{reconst2plus1}
  r\colon(A_a^i\tau_i\md x^a,\phi^i\tau_i)\mapsto A^i_a\tau_i\md x^a+
   \phi^i\tau_i\md z.
\end{equation}
Here the matrices $\tau_i=-\frac{i}{2}\sigma_i$ form a basis of
$\Lie G$ and $x^a$, $a=1,2$ are coordinates on a chart of $B$.
This shows that the Higgs field gives the $z$-component of the
connection form.

Following the steps of the last section, we obtain the
configuration space $\AbR\times\UbR$ of the reduced theory which
consists of the fields $A_a^i$ and $\phi^i$ on $B$ appearing in
(\ref{reconst2plus1}).

Up to now we have imposed only the symmetry condition, not the
condition on the norm of the Killing field. This second condition
serves to remove the dynamical scalar field. An equivalent quantum
condition can be formulated by allowing only trivial Higgs
vertices.  This will eliminate the dependence of the spin networks
on generalized Higgs fields, thereby removing their local degrees
of freedom. So we arrive at the quantum configuration space $\AbR$
of vacuum $2+1$ dimensional gravity.

We can now build an auxiliary Hilbert space of integrable
cylindrical functions on $\AbR$, completely analogous to the
unreduced theory. On this Hilbert space which is spanned by
ordinary two-dimensional spin networks we can represent the
constraints and search for solutions. This has already been done
in Ref.\ \cite{QSDIV} starting from the classical symmetry
reduction and quantizing the classical configuration space. Our
quantum symmetry reduction procedure gives the same results, but
with a representation of the quantum states of the reduced theory
as symmetric states of the full quantum theory.

The reduction of this subsection can be generalized
straightforwardly  to the case of an arbitrary gauge theory on
$\Sigma=B\times\R^n$, where $B$ is an arbitrary $d$-dimensional
manifold and $S=\R^n$ acts on the second factor of $\Sigma$. A
symmetry reduction will involve $n$ components of a Higgs field,
whereas a {\em quantum dimensional
  reduction} from $B\times\R^n$ to $B$ can be obtained by
  postulating
trivial Higgs vertices of spin networks in $B$.

\subsection{Spherically Symmetric Electromagnetism}
\label{s:SphSymmElec}

We will now reduce electromagnetism to a spherically symmetric
one. Although spherically symmetric electromagnetism is almost
trivial it is nevertheless quite instructive for our purposes
because the quantum symmetry reduction can be carried out
explicitly.

 First, let us explain why we treat electromagnetism as a
diffeomorphism invariant theory. This can, of course, not be true
for pure electromagnetism on, e.g.\ a Minkowski background. But we
are interested in the electromagnetic field as a field of a
Reissner-Nordstr{\o}m black hole.  We therefore have to couple
electromagnetism to gravity, which ensures diffeomorphism
invariance. The dynamics of the electromagnetic field is then
encoded in the Hamiltonian constraint. The diffeomorphism
constraint will only contain gravitational fields, because in the
spherically symmetric case diffeomorphism invariance of the
electromagnetic field is already imposed by the $U(1)$-Gau{\ss}
constraint.  As a test model this coupled theory as a whole would
be too complicated, so we discard the gravitational field by
constraining it to be zero. This amounts to treating the
electromagnetic field on a degenerate background which renders the
Hamiltonian ill-defined. But electric and magnetic fluxes will be
well-defined, so that we, nevertheless, may study the kinematics
of the fields.

\subsubsection{Classical Symmetry Reduction}

We now have $G=U(1)$, $S=SU(2)$ and in general $F=U(1)$. The
topology of space is $\Sigma\cong\R\times S^2$ or
$\Sigma\cong\R^+\times S^2$, implying $B=\R$ or $B=\R^+$,
respectively.

 We first determine all conjugacy classes of the homomorphisms
$\lambda\colon U(1)\to U(1)$. Because such a homomorphism is a
one-dimensional unitary representation of $U(1)$, it is given by
its character. Therefore we have the mappings
$\Hom(U(1),U(1))/\Ad\cong\Z$ represented by the homomorphisms
$\lambda_n\colon z\mapsto z^n$. For the abelian group $G$ the
centralizers $Z_{\lambda_n}=U(1)$ are equal to the full group.
Thus, a spherically symmetric principal fibre bundle with
structure group $U(1)$ is classified by an integer $n$, and, of
course, all reduced bundles are trivial, because $B$ is
contractible, and they are not needed for the classification.

Let us now determine the type of Higgs field allowed by Eq.\
(\ref{Higgs}). This field is given by the map $\Lambda\colon\Lie
S\to\Lie G$, where $\Lambda|_{\Lie F}=\md\lambda_n$ is already
fixed by the bundle classification. Here we have $\dim\Lie G=1$,
leaving only the possibilities $\dim\Kern\Lambda=2$ or
$\dim\Kern\Lambda=3$. In the case $n\not=0$, $\md\lambda_n$ is an
isomorphism of vector spaces, forbidding $\dim\Kern\Lambda=3$. So
we have $\dim\Kern\Lambda=2$ and the Higgs field
$\phi=\Lambda|_{\Lie F_{\perp}}=0$ vanishes. This kind of
reasoning can not be used in the case $n=0$. But looking at Eq.\
(\ref{Higgs}) we see that $n$ enters this equation only in
connection with $\mbox{Ad}_{\lambda_n(f)}$ for $f\in F$. This
occurrence is trivial in the abelian case, so Eq.\ (\ref{Higgs})
is independent of $n$. Therefore, if it does not allow a Higgs
field for $n\not=0$ it cannot allow a Higgs field for $n=0$, too.
This proves that the reduced theory is a theory of a
$U(1)$-connection only.

Let us recall the reduced phase space structure from Ref.\
\cite{thi2}. The canonical variables of the symmetry reduced
theory are given by the radial components $p$ and $i\,\omega$ of
the electric field and the $U(1)$-connection form, respectively.
They are subject to the Gau\ss\ constraint $p^{\prime}\approx 0$
which enforces $p$ to be constant. The reduced phase space is
two-dimensional with the canonical variables $p$ which is
proportional to the electric charge and the canonically conjugate
$\Phi=-\int_B\md x\,\omega$.  The electric charge is either a
point charge sitting in $x=0$ if $B=\R^+$, or the charge of the
wormhole $\Sigma=\R\times S^2$.

\subsubsection{Quantum Symmetry Reduction}

According to Theorem~\ref{ident} the physical Hilbert space of
spherically symmetric electromagnetism is given by
$L_2(\AbR/\GbR,\md\mu_{AL})$, the space of gauge invariant,
Ashtekar-Lewandowski square integrable functions on the space of
generalized $U(1)$-connections over $B$.  It is spanned by
$U(1)$-spin networks, which are, due to gauge invariance and the
one-dimensional nature of $B$, given by
\begin{equation}\label{etaBK}
  (\eta_B)^K\quad , \quad\mbox{where }\eta_B:=\exp i\int_B\md x\,\omega(x)
\end{equation}
is the holonomy along $B$ of the $U(1)$-connection $i\,\omega$ and
$K\in\Z$ is the only parameter, corresponding to irreducible
$U(1)$-representations labeling the basis states. We see that the
reduced theory is particularly simple, having only two canonical
degrees of freedom. The physical Hilbert space can be identified
with $L_2(U(1),\md\mu_H)$, where $\md\mu_H$ is the Haar measure on
$U(1)$.

Before coming to the observables of the theory we will investigate
the quantum symmetry reduction $\rstar$. Let the symmetric
principal fibre bundle be classified by the homomorphism
$\lambda_n$, and let $F=U(1)\cong\exp\langle\tau_3\rangle\subset
SU(2)$ be the isotropy subgroup. We then have to set
$\Lambda(\tau_3)=\md\lambda_n(\tau_3)=i\,n$ and
$\Lambda(\tau_{1,2})=0$ in Eq.\ (\ref{ASF}).  The reconstruction
of a $U(1)$-connection with $x$-component $i\,\omega_x\colon B\to
i\, \R=\Lie U(1)$ on $Q$ is given by the $[\lambda_n]$-invariant
connection
\begin{equation}\label{omegaU1}
  i\,\omega:=r_{\lambda_n}^{(\iota)}(i\,\omega_x\, \md x)=i\,\omega_x\,\md x+
   i\,n\cos\vartheta\, \md\varphi.
\end{equation}
Here $x$ is a (local) coordinate of $B$ and $(\vartheta,\varphi)$
are the Killing parameters of the $S$-orbits. Together they yield
a (local) coordinate system $(x,\kt,\kp)$ of $B\times S^2$.  We
can calculate the curvature of this $[\lambda_n]$-invariant
connection to obtain the (densitized) magnetic field of a Dirac
monopole with magnetic charge $n$:
\begin{equation}
  B_x=-n\sin\vartheta\quad , \quad B_{\vartheta}=B_{\varphi}=0.
\end{equation}
This confirms our claim that the conjugacy classes of
homomorphisms correspond to the values of a topological charge.

Eq.\ (\ref{omegaU1}) will now be used to pull back a spin network
state to a function on the space of invariant connections. To that
end let $e\colon[0,1]\to\Sigma$ be an edge in $\Sigma$, running
from $e(0)=(x_1,\kt_1,\kp_1)$ to $e(1)=(x_2,\kt_2,\kp_2)$, chosen
such that a parameterization $\kt(\kp)$ is possible along $e$
(otherwise we can cut $e$ in pieces and set $\kt(\kp)=0$ if $\kp$
is constant along a piece without affecting the following). Then
the $i\,\omega$-holonomy along $e$ is given by
\begin{eqnarray*}
 h_e(i\omega) & = & \exp[\int_e\md t(i\, \dot{x}\,\omega_x+
  i\,n\,\dot{\kp}\cos\kt)]\\ &=& \exp\left(i\int_{x_1}^{x_2}
  \md x\omega_x\right)\exp\left(in\int_{\kp_1}^{\kp_2}
  \md\kp\cos\kt(\kp)\right)\\
 & =: & h_{\pi(e)}(i\omega_x)(\beta_e)^n~.
\end{eqnarray*}
Here $\pi(e):=[x_1,x_2]=[x(e(0)),x(e(1))]\subset B$ is an edge in
$B$, and $h_{\pi(e)}(i\,\omega_x)$ is the holonomy of the reduced
connection $i\,\omega_x$ along it. $\beta_e$ is a phase factor
depending only on the geometry of $e$.

If $T_{\gamma,\,k}$ is a spin network with a graph
$\gamma\subset\Sigma$ and a labeling $k$ of its edges with
irreducible $U(1)$-representations (a labeling of the vertices is
not necessary for gauge invariant $U(1)$-spin networks because it
is already given uniquely by the edge labeling), we can evaluate
it on a $[\lambda_n]$-invariant connection: If $\Eg$ denotes the
edge set of $\gamma$ we have
\begin{eqnarray}\label{Tgk}
  T_{\gamma,\, k}(i\omega)&=&\prod_{e\in\Eg}h_e(i\omega)^{k_e}
  \\ &=&\prod_{e\in\Eg}h_{\pi(e)}(i\omega_x)^{k_e}\prod_{e\in\Eg}
  (\beta_e)^{nk_e}=(\beta_{\gamma,\, k})^n
  \prod_{e\in\Eg}h_{\pi(e)}(i\omega_x)^{k_e}. \nonumber
\end{eqnarray}
Here $\beta_{\gamma,\, k}:=\prod_{e\in\Eg}\beta_e^{k_e}$ is a
phase factor depending only on the geometry of $T$ and its
labeling.

The right hand side of the last equation can be written as the
evaluation of a spin network state in $B$ on the connection
$i\,\omega_x$. In order to do this we will define a projection
$\pi$ which assigns to a spin network $T_{\gamma,\, k}$ in
$\Sigma$ a spin network $\pi(T_{\gamma,\, k})$ in $B$. The set of
vertices $\Vg$ of $\gamma$ can be projected on a finite set
$$\pi(\Vg):=\{x(v):v\in\Vg\}=:\{x^{(i)}\}_{1\leq
i\leq|\pi(\Vg)|}~,$$ where we have ordered the elements of
$\pi(\Vg)$ such that $x^{(i)} < x^{(j)}$ for $i<j$. Their number
is bounded by $|\pi(\Vg)|\leq|\Vg|$. Let $\gamma$ be chosen --~if
necessary by inverting edges and splitting edges by introducing
new vertices~-- such that each edge $e\in\Eg$ either lies entirely
in an orbit of $S$, in which case we define $\pi(e):=\emptyset$,
or it has a projection $\pi(e)=[x^{(i)},x^{(i+1)}]$  for some $1
\leq i \leq |\pi(V(\gamma)|$ and $x(e(t))$ increases monotonely in
$t$. To every projected edge $\pi(e)$ we assign the point
$x_m(\pi(e)):=\frac{1}{2}(x(e(0))+x(e(1)))$ in its interior. We
can now define the projected spin network:

\begin{defi}\label{project}
  Let $T_{\gamma,\, k}$ be a $U(1)$-spin network in $\Sigma=B\times S^2$,
  the graph $\gamma$ be chosen as above.

  The {\em projected graph} $\pi(\gamma)\subset B$ is given by its edge set
$$  E(\pi(\gamma)):=\{\pi(e):e\in\Eg,\pi(e)\not=\emptyset\}=
 \{[x^{(i)},x^{(i+1)}]:1\leq i<|\pi(\Vg)|\}
$$
  and its vertex set
$$  V(\pi(\gamma)):=\pi(\Vg). $$

  The labeling of the projected graph descending from the spin network
  $T_{\gamma,\, k}$ is given by
$$  \pi(k)_{\pi(e)}:=\sum_{|e^{\prime}\cap S_{x_m(\pi(e))}|=1}
  k_{e^{\prime}}
$$
  for any edge $\pi(e)\in E(\pi(\gamma))$. Here $S_{x_m(\pi(e))}$ is
  the $S$-orbit through $x_m(\pi(e))$, and the condition
  $|e^{\prime}\cap S_{x_m(\pi(e))}|=1$ in the sum ensures that we
  count only the charge of edges running transversally through
  $S_{x_m(\pi(e))}$.

  The {\em projected spin network} is given by
  $\pi(T_{\gamma,\, k}):=T_{\pi(\gamma),\pi(k)}$.
\end{defi}

By means of the projected spin network, we can write Eq.\
(\ref{Tgk}) as
\begin{eqnarray*}
 T_{\gamma,\, k}(i\omega) & = &(\beta_{\gamma,\, k})^n
  \prod_{e_B\in E(\pi(\gamma))}\;\;
  \prod_{e^{\prime}\in\Eg:\pi(e^{\prime})=e_B}h_{e_B}
  (i\omega_x)^{k_{e^{\prime}}}\\
 & = &(\beta_{\gamma,\, k})^n\prod_{e_B\in E(\pi(\gamma))}h_{e_B}
  (i\omega_x)^{\pi(k)_{e_B}}=
  (\beta_{\gamma,\, k})^n \pi(T_{\gamma,\, k})(i\omega_x).
\end{eqnarray*}
This equation enables us to write
\begin{equation}
  \rnstar T_{\gamma,\, k}=(\beta_{\gamma,\, k})^n \pi(T_{\gamma,\, k}).
\end{equation}
We can see that $\rnstar T\in\Phi_B$ is cylindrical on $\AbR$ for
any spin network $T\in\Phi_{\Sigma}$.  This convenient
circumstance is related to the vanishing of the Higgs field and
cannot be taken for granted in cases of other structure groups.
Here this fact allows us to compose the maps $\rho_{[\lambda]}$
and $\sigma_{[\lambda]}$ in the diagram at the end of
Section~\ref{s:construct} to obtain a map
$$
  \sigma_{[\lambda]}\circ\rho_{[\lambda]}\colon
  \Phi_{\Sigma}\to\Phi_{\Sigma}^{\prime}
$$
reminiscent of a group averaging map. In fact the symmetry
reduction considered here can be imposed by supplementing
electrodynamics with a further constraint besides the Gau\ss\
constraint to arrive at an abelian $BF$-theory \cite{BFElec}.
Thus, the symmetry can be reduced by means of a ``rigging'' map
analogous to $\sigma_{[\lambda]}\circ\rho_{[\lambda]}$. The
vanishing Higgs field also causes the appearance of the
topological charge $n$  as a power of the phase factor only.
Furthermore, the above calculations exhibit the fact that a
projected spin network state is gauge invariant if and only if the
original spin network state in $\Sigma$ is gauge invariant. This
follows from the fact that gauge invariance in $\Sigma$ forces the
spin network labeling to fulfill
$$
 \sum_{|e\cap S_1|=1}k_e=\sum_{|e\cap S_2|=1}k_e
$$
for any two $S$-orbits $S_1$ and $S_2$ not containing a vertex of
the graph $\gamma$ (for simplicity we assume that all edges are
oriented outwards). Therefore a gauge invariant spin network will
project to a spin network with labeling $k_{e_B}=K$ for each edge
$e_B$ of the projected graph. This yields the gauge invariant spin
network $(\eta_B)^K$ defined above as projected spin network.

Let us now write down the action of a $[\lambda_n]$-symmetric
state as distribution on $\Phi_{\Sigma}$.  For a spin network
$T_{\gamma,\, k}$ it is given by
\begin{eqnarray}
  \sigln((\eta_B)^K)(T_{\gamma,\, k}) & = & \int_{\AbR}\md\mu_{AL}\:
   (\overline{\eta}_B)^K\:\rnstar T_{\gamma,\, k}\\
   &=&(\beta_{\gamma,\, k})^n\int_{\AbR}\md\mu_{AL}\:(\overline{\eta}_B)^K \:
   T_{\pi(\gamma),\, \pi(k)}\nonumber\\
  & = & (\beta_{\gamma,\, k})^n\:\delta_{\{B\},\,\pi(\gamma)}\:
  \delta_{\{K\},\,\pi(k)}~. \nonumber
\end{eqnarray}
It is non-vanishing only if $T_{\gamma,\, k}$ is gauge invariant
implied by $\pi(\gamma)=\{B\}$, and if the charge
$\pi(k)=\sum_{|e\cap S_x|=1}k_e$ equals the labeling $K$ of the
symmetric state.

\subsubsection{Observables}

We will now quantize the observables $p$ and $\Phi$ found in the
classical reduction. $p$ is canonically conjugate to the
connection $\omega$, and standard quantization rules lead in the
connection representation to the quantization
$\hat{p}=-i\hbar\fd{\omega}$. Acting on a gauge invariant state
(\ref{etaBK}) this yields $\hat{p}\:(\eta_B)^K=\hbar K(\eta_B)^K$.
We see that spin network states are eigenstates of $\hat{p}$. All
the eigenvalues of $\hat{p}$ are real, which shows that it is
essentially selfadjoint. Furthermore, they are discrete exhibiting
electric charge quantization as integer multiples of an elementary
charge (this fact has already been observed in Ref.\
\cite{MaxwellSpinnet} in the unsymmetric theory). The value of
this elementary charge is, however, not fixed because $p$ is only
proportional to the electric charge. The factor of proportionality
 ensures the correct dimension and scales the elementary charge.
This factor is arbitrary and  is related to the normalization of
the electromagnetic action.

In the following we will denote the basis states of the Hilbert
space $L_2(U(1),\md\mu_H)$ as
$$  \ket{K}:=(\eta_B)^K.   $$
These states span an orthonormal basis with respect to the Haar
measure on $U(1)=\AbR/\GbR$ which derives from the
Ashtekar-Lewandowski measure on $\AbR$. They are labeled by the
charge eigenvalues according to $\hat{p}\ket{K}=\hbar K\ket{K}$.

Now we have to quantize $\Phi=-\int_B\omega=i\log\eta_B$. Recall
that holonomies are well-defined as operators in the loop
quantization. This fact suggests to use $\eta_B$ instead of
$\Phi$. $\eta_B$ can straightforwardly be promoted to the
`creation' operator
$$  \hat{\eta}_B\ket{K}=(\eta_B)^{K+1}=\ket{K+1},   $$
which is unitary because it is invertible and preserves the norm
of states it acts on. In view of $\eta_B=\exp(-i\Phi)$ the
operator  $\hat{\eta}_B$ has indeed to be unitary. Note that it is
the Haar measure on $U(1)$, which derives from the
Ashtekar-Lewandowski measure on the unconstrained space (before
solving the Gau\ss\ constraint) that incorporates the classical
reality conditions correctly.

We will now close this subsection by showing that $\hat{p}$ and
$\hat{\eta}_B$ have the correct commutator.

\begin{theo}
  $(p,\exp(-i\Phi))\mapsto(\hat{p},\hat{\eta}_B)$ is a representation
  of the classical Poisson $\star$-algebra on the Hilbert space
  $L_2(U(1),\md\mu_H)$.
\end{theo}

\begin{proof}
  We have already seen, that the $\star$-relations $p^{\star}=p$ and
  $\exp(-i\Phi)^{\star} =\exp(i\Phi)$ are represented properly.

  Because $p$ and $\Phi$ are canonically conjugate, we have the Poisson bracket
$$  \{p,\exp(-i\Phi)\}=\td[\exp(-i\Phi)]{\Phi}=-i\exp(-i\Phi).  $$
  The only non-vanishing matrix elements of the commutator
  $[\hat{p},\hat{\eta}_B]$ are given by
$$
  \DirB[\mbox{$[\hat{p},\hat{\eta}_B]$}]{K}{K-1}=\DirB[\hbar K]{K}{K}-
   \DirB[\hbar(K-1)]{K}{K}=\hbar.
$$
  We, therefore, have the relation
 \begin{equation}
  [\hat{p},\hat{\eta}_B]=\hbar\hat{\eta}_B=i\hbar\{p,\exp(-i\Phi)\}\hat{\:}
 \end{equation}
  implementing the correct representation of the Poisson structure.
\end{proof}

\subsection{Spherically Symmetric Quantum Gravity}

Our last example deals with spherically symmetric quantum gravity.
Undoubtedly this one of our three examples is of the greatest
physical interest because it describes quantum properties of
non-rotating black holes. The classical symmetry and constraint
reduction reveals that there is only one physical configuration
degree of freedom, the mass (or a canonical pair: mass and time).
This is similar to spherically symmetric electromagnetism
discussed above which has only the electric charge as physical
degree of freedom. However, it can be anticipated that gravity
will be more difficult since its constraints are more complicated.

In the present paper we treat in detail only the Gau{\ss}
constraint and comment briefly on the solution of the
diffeomorphism constraint. The more complicated Hamiltonian
constraint will be dealt with elsewhere \cite{bo3}.

\subsubsection{Symmetry Reduction}

We now specialize our general framework to the case $S=SU(2)$,
$F=U(1)$ and $G=SU(2)$ (we will use real Ashtekar variables). The
topology of the space $\Sigma$  will be $\Sigma=B\times S^2$ with
$B=\R$ or $B=\R^+$.

 First we have to find all conjugacy classes of homomorphisms
$\lambda\colon F=U(1)\to SU(2)=G$.  In order to do that we can
make use of Eq.\ (\ref{Hom}). We need the following information
about $SU(2)$ (see, e.g.\ Ref.\ \cite{BroeckerDieck}).

\begin{lemma}
  The {\em standard maximal torus} of $SU(2)$ is given by
$$  T(SU(2))=\{\diag(z,z^{-1})|\, z\in U(1)\}\cong U(1).  $$
  The {\em Weyl group} of $SU(2)$ is the permutation group of two
  elements, $$W(SU(2))\cong S_2~,$$ its generator acting on $T(SU(2))$ by
  $\diag(z,z^{-1})\mapsto\diag(z^{-1},z)$.
\end{lemma}

All homomorphisms in $\Hom(U(1),T(SU(2)))$ are then given by
$$\lambda_k\colon z\mapsto\diag(z^k,z^{-k})$$ for any $k\in\Z$, as
in the electromagnetic example. However, here we have to divide
out the action of the Weyl group, leaving only the maps
$\lambda_k$, $k\in\N_0$, as representatives of all conjugacy
classes of homomorphisms. We see that spherically symmetric
gravity has a topological charge taking values in $\N_0$ (in the
dreibein, not in the metric formulation, as we will see below).

We will represent $F$ as the subgroup
$\exp\langle\tau_3\rangle<SU(2)$ of the symmetry group $S$, and
use the homomorphisms $\lambda_k\colon\exp t\tau_3\mapsto\exp
kt\tau_3$, out of each conjugacy class. This amounts to a partial
gauge fixing called $\tau_3$-gauge in the following.  As reduced
structure group we obtain
$Z_G(\lambda_k(F))=\exp\langle\tau_3\rangle\cong U(1)$ for
$k\not=0$ and $Z_G(\lambda_0(F))=SU(2)$ ($k=0$ leads to the
manifestly symmetric connections of Ref.\ \cite{CorderoTeit}). The
map $\Lambda|_{\Lie F}$ is then given by
$\md\lambda_k\colon\langle\tau_3\rangle\to\Lie G,\tau_3\mapsto
k\tau_3$. The remaining components of $\Lambda$ which give us the
Higgs field, are determined by $\Lambda(\tau_{1,\,2})\in\Lie G$,
and are subject to Eq.\ (\ref{Higgs}). This equation can here be
written as
$$\Lambda\circ\ad_{\tau_3}=\ad_{\md\lambda(\tau_3)}\circ\Lambda~.$$
Using $\ad_{\tau_3}\tau_1=\tau_2$ and $\ad_{\tau_3}\tau_2=-\tau_1$
we obtain the equation
$$   \Lambda(a_0\tau_2-b_0\tau_1)=k(a_0[\tau_3,\Lambda(\tau_1)]+
  b_0[\tau_3,\Lambda(\tau_2)]),
$$
where $a_0\tau_1+b_0\tau_2$, $a_0,b_0\in\R$ is an arbitrary
element of $\Lie F_{\perp}$.  Since $a_0$ and $b_0$ are arbitrary
this is equivalent to the two equations
$$  k[\tau_3,\Lambda(\tau_1)]=\Lambda(\tau_2)\quad\mbox{and}\quad
    k[\tau_3,\Lambda(\tau_2)]=-\Lambda(\tau_1).
$$
The general ansatz
$$   \Lambda(\tau_1)=a_1\tau_1+b_1\tau_2+c_1\tau_3\quad,\quad
     \Lambda(\tau_2)=a_2\tau_1+b_2\tau_2+c_2\tau_3
$$
with arbitrary parameters $a_i,b_i,c_i\in\R$ yields the equations
\begin{eqnarray}
k(a_1\tau_2-b_1\tau_1)&=& a_2\tau_1+b_2\tau_2+c_2\tau_3~, \\
     k(-a_2\tau_2+b_2\tau_1)& =& a_1\tau_1+b_1\tau_2+c_1\tau_3~.
     \nonumber
\end{eqnarray}
These equations have a non-trivial solution only if $k=1$, for
which we get $$b_2=a_1~,~~b_2 =-b_1~~\mbox{ and }~~  c_1=c_2=0~.$$
We shall discuss this main case first where we also will see how
the present approach is related to that of Refs.
\cite{thika,kathi} and will comment on the physically
uninteresting ones ($k \neq 1$) which have a vanishing Higgs field
afterwards.

 The configuration variables of the system are the above
 fields $a,b,c~\colon B\to\R$
of the  $U(1)$-connection form $A=c(x)\,\tau_3\,\md x$ on the one
hand and the two Higgs field components
\begin{eqnarray}
 \Lambda|_{\langle\tau_1\rangle}\colon B&\to&\langle\tau_1,\tau_2\rangle,
~  x\mapsto a(x)\tau_1+b(x)\tau_2\\ & = & \frac{1}{2}\left(
  \begin{array}{cc} 0 & -b(x)-ia(x)\\
  b(x)-ia(x) & 0\end{array}\right)
 =\colon \left(\begin{array}{cc} 0 & -\overline{w}(x)\\w(x) & 0
  \end{array}\right) \nonumber
\end{eqnarray}
on the other hand.

Under a local $U(1)$-gauge transformation $z(x)=\exp t(x)\tau_3$
they transform as $$ c\mapsto c+\td[t]{x}~ \mbox{ and }
w(x)\mapsto\exp(-it)w$$ which can be read off from
\begin{eqnarray*} A& \mapsto &z^{-1}Az+z^{-1}\md z=A+\tau_3\md t~,
 \\ \Lambda(\tau_1)&\mapsto & z^{-1}\Lambda(\tau_1)z=\left(
    \begin{array}{cc} 0 & -\exp(it)\overline{w}\\ \exp(-it)w & 0
    \end{array}\right).
\end{eqnarray*}
Comparing with Ref.\ \cite{thika} we see that the above variable
$c$ transforms as the connection coefficient $A_1$ there and the
variables $(a,b)$ as $(A_2,A_3)$. However, the reconstructed
connection form
\begin{eqnarray} A(x,\kt,\kp)&=& c(x)\tau_3\md
x+(-a(x)\md\kt+b(x)\sin\kt\md\kp)
  \tau_1-\nonumber \\ && -(b(x)\md\kt
  +a(x)\sin\kt\md\kp)\tau_2+\cos\kt\,\md\kp\, \tau_3
\end{eqnarray}
is different from the connection form
\begin{eqnarray}&& \left(A_1n^i_x\md x+2^{-\frac{1}{2}}(A_2n^i_{\kt}+
   (A_3-\sqrt{2})n^i_{\kp})\md\kt+\right. \nonumber \\
   && +2^{-\frac{1}{2}}(A_2n^i_{\kp}-\left.(A_3-\sqrt{2})n^i_{\kt})
   \sin\kt\md\kp\right)\tau_i
\end{eqnarray}
used in Refs.\ \cite{thika,kathi} (now expressed in terms of real
Ashtekar variables) where $n_x^i$, $n_{\kt}^i$ and $n_{\kp}^i$ are
the standard unit vectors in polar coordinates. The two connection
forms differ, however, only by a gauge rotation $g_1:=\exp(-\kt
n_{\kp}^i\tau_i)$ followed by a second gauge rotation
$g_2:=\exp(-\kp n_x^i\tau_i)$.

We note that the term $\cos\kt\, \md\kp\, \tau_3$ in Eq.\ (25)
which is the only term independent of the fields $a$, $b$ and $c$,
in the connection form $A(x,\kt,\kp)$ derived above leads
automatically to the appearance of $A_3$  as the combination
$A_3-\sqrt{2}$ in the connection form (26) of Refs.\
\cite{thika,kathi}. This subtraction of $\sqrt{2}$ (which does not
appear in Ref.\ \cite{CorderoTeit}) has been added by hand in
Ref.\ \cite{thika} in order to ensure the correct transformation
of $(A_2,A_3)$ in the defining representation of $SO(2)$.

The calculation above shows that $c$ has to be identified with
$A_1$ and (up to a rescaling with $\sqrt{2}$ used in Ref.\
\cite{thika} in order to get the standard symplectic structure)
$(a,b)$ with $(A_2,A_3)$. We will here rescale the variables, too,
and denote the rescaled ones by $(A_1,A_2,A_3)$. Their conjugate
variables are $(E^1,E^2,E^3)$, and the symplectic structure is
--~adapted to our notation using the Immirzi parameter $\iota$~--
given by
\begin{equation}\label{sympl}
  \{A_I(x),E^J(y)\}=\frac{\kappa\iota}{4\pi}\delta^J_I\delta(x,y)
\end{equation}
where $\kappa=8\pi G$ is the gravitational constant.

We recall \cite{thika,kathi} that the metric on $\Sigma$ is given
by \begin{equation}
(q_{ab})=\mbox{diag}(\frac{E}{2E^1},E^1,E^1\,\sin^2\vartheta)~,~
E=(E^2)^2+(E^3)^2~. \end{equation} The functions $E^1$ and $E$ are
closely related to two important geometrical quantities:

The surface $A(x)$ of the 2-dimensional spherical orbit generated
by the symmetry group $S$ at $x \in B$ is given by
\begin{equation} A(x)=4\pi\, E^1(x)\end{equation} and the
spherically symmetric three-dimensional volume element $dV$
``transverse'' to those orbits is \begin{equation} dV =
\frac{4\pi}{\sqrt{2}}\sqrt{E^1\,E}\;dx ~.\end{equation}

We finally turn briefly to  the case $k\not=1$ which is associated
with vanishing Higgs fields $\Lambda(\tau_i)=0, i=1,2$ or,
equivalently, $A_2=A_3=0$. The vanishing of the canonical
variables $A_2$ and $A_3$ implies that their canonically conjugate
momenta $E^2$ and $E^3$ become irrelevant, too, and we may put
them to zero, $E^2=E^3=0$. This means that the metric (28) and the
volume (30) become degenerate and that we do not have a
non-vanishing volume element.

Thus, the sectors with $k\not=1$ describe degenerate sectors which
are, however, different from the one found in Ref.\ \cite{thika}.
There a degenerate sector with vanishing volume was found in our
($k=1$)-sector after solving the constraints. On our
($k\not=1$)-sectors the diffeomorphism and Hamiltonian constraint
will be trivially fulfilled, but the $k=1$-sector has nontrivial
constraints whose solution manifold has several sectors.

That the degenerate sectors for $k\neq 1$ here are different from
the degenerate one of Ref. \cite{thika} can be seen as follows:
For $k \neq 1$ the last term ($\cos \vartheta\, d\varphi\,
\tau_3$) in Eq.\ (25) gets multiplied by $k$ and as a consequence
the subtracted constant $-\sqrt{2}$ in Eq.\ (26), too, leading to
different connection forms even for $A_2=A_3=0$.

Thus, if we are interested in geometrically interesting systems
with non-vanishing volumes we have to stick to the sector $k=1$.

\subsubsection{Quantization of the Gau{\ss} Constraint}\label{Gauss}

Although the Gau{\ss}  constraint can be solved by restricting to
gauge invariant spin networks we here give  a regularization in
order to exhibit more concretely the role of the reduced gauge
group.

Classically, one uses a partial gauge fixing reducing the
structure group $SU(2)$ to $U(1)$.  This is explicit in the
symmetry reduction of the Gau\ss\ constraint
$$   {\cal G}^E[\Lambda]=\frac{4\pi}{\kappa\iota}\int_B\md x
      \Lambda(x)\left((E^1)^{\prime}+A_2E^3-A_3E^2\right)
$$
which is taken from \cite{thika} adapted to our notation.
$\Lambda(x)$ is a Lagrange multiplier.  In a pure, one-dimensional
$U(1)$-gauge theory one would encounter the constraint
$(E^1)^{\prime}\approx0$.  Here, we have an additional term which
provides coupling to the Higgs field. For the sake of clarity we
will regularize these two terms one after another.

By standard methods we get
\begin{eqnarray*}
 && \frac{4\pi}{\kappa\iota}\int_B\md x\Lambda(\hat{E}^1)^{\prime}
   f_{\gamma}  =\\ && =
   \frac{\hbar}{i}\int_B\md x\Lambda\td{x}\sum_e\int_e\md t
   \dot{e}^x\delta(x,e(t))\tr\left((h_e(0,t)\tau_3h_e(t,1))^T
   \pd{h_e}\right)f_{\gamma}\\
 & &=  i\hbar\sum_e\int_e\md t\dot{e}^x\int_B\md x\td[\Lambda]{x}\delta(x,e(t))
   \tr\left((h_e(0,t)\tau_3h_e(t,1))^T\pd{h_e}\right)f_{\gamma}\\
 & &=  i\hbar\sum_e\int_e\md t\td{t}\left(\Lambda(e(t))
   \tr\left((h_e(0,t)\tau_3h_e(t,1))^T\pd{h_e}\right)\right)f_{\gamma}\\
 & &=  i\hbar\sum_e\left(\Lambda(e(1))X_L^3(h_e)-\Lambda(e(0))
   X_R^3(h_e)\right)f_{\gamma}.
\end{eqnarray*}
In the first row, we used the standard quantization rule
$$\hat{E}^1=\frac{\kappa\hbar\iota}{4\pi i}\fd{A_1}$$ taking into
account the symplectic structure (\ref{sympl}). The functional
derivative acts on a cylindrical function $f_{\gamma}$ which
depends on edge and point holonomies. As remarked in
Section~\ref{SymmState} we start the quantization procedure on the
space $(\overline{\cal A}\times\overline{\cal U})^{\lambda}$ due
to classical gauge fixing. With  $\lambda$ chosen as in the
present section a holonomy on this space takes the form
$$h_e=\exp(\int_e\md x A_1(x)\tau_3)$$ giving rise to the above
derivative.

In the second step above we integrated by parts to be able to
integrate over the $\delta$-distribution in the following step.
Because there the holonomies are abelian, the trace does actually
not depend on $t$ so that we can include it into the argument of
the $t$-derivative in order to integrate over $t$. Because of the
abelian nature the left and right invariant vector field
components $X_L^3$ and $X_R^3$ are identical. However, we keep
both of them to make possible a comparison with a $SU(2)$-theory.

The Higgs coupling term has to be rephrased before proceeding: We
introduce polar coordinates $(A,\alpha)$ in the $(A_2,A_3)$-plane,
i.~e., $(A_2,A_3)=(A\cos\alpha,A\sin\alpha)$. These variables have
the advantage that $A$ is gauge invariant, whereas $\alpha$ can be
gauged arbitrarily.  After replacing the variables $E^2$ and $E^3$
with the respective derivative operators in the course of
quantization, we encounter the derivation
$$   A_2\pd{A_3}-A_3\pd{A_2}=\pd{\alpha}  $$
the action of which on a point holonomy $h_v(A,\alpha)=\exp
A(\cos\alpha\tau_1+\sin\alpha\tau_2)$ (in our $\tau_3$-gauge) can
be calculated, using
$$  h_v(A,\alpha)=\exp(\alpha\tau_3)h_v(A,0)\exp(-\alpha\tau_3)~,
$$
to be
$$   \pd[h_v]{\alpha}=\tau_3h_v-h_v\tau_3.  $$
Now we are in a position to quantize the remaining part of the
Gau\ss\ constraint:
\begin{eqnarray*}
  &&\frac{4\pi}{\kappa\iota}\int_B\md x\Lambda(x)(A_2E^3-A_3E^2)
    \,\hat{}\:f_{\gamma}   \\&& =
    \frac{\hbar}{i}\int_B\md x\Lambda(x)\fd{\alpha(x)}f_{\gamma} \\
  && =  \frac{\hbar}{i}\int_B\md x\Lambda(x)\sum_{v\in B}\delta(x,v)
    \tr\left((\tau_3h_v-h_v\tau_3)^T\pd{h_v}\right)f_{\gamma}\\
  & &=  i\hbar\sum_{v\in B}\Lambda(v)\left(X_L^3(h_v)-
    X_R^3(h_v)\right)f_{\gamma}~.
\end{eqnarray*}
We can now combine the operators to obtain the quantization
\begin{equation}
   \hat{{\cal G}}^E[\Lambda]=i\hbar\sum_{v\in B}\Lambda(v)\left(\sum_{e(1)=v}
  X_L^3(h_e)-\sum_{e(0)=v}X_R^3(h_e)+X_L^3(h_v)-X_R^3(h_v)\right)
\end{equation}
of the Gau\ss\ constraint which has a finite action on cylindrical
functions and is therefore densely defined.

The structure of the operator is as expected: It is a sum over
vertices weighted with the Lagrange multiplier. In each vertex it
acts as the sum of a left invariant vector field for each incoming
edge, a right invariant vector field for each outgoing edge, and a
left as well as right invariant vector field for the point
holonomy in the vertex. That the point holonomy contributes by
left and right invariant vector fields represents the fact that
the Higgs field transforms in the adjoint representation of
$SU(2)$. The sum of vector fields is in complete agreement with
the definition of gauge invariant spin networks with Higgs.
However, $\hat{{\cal G}}^E$ is a sum only of the third components
of the vector fields in accordance with our $\tau_3$-gauge. Recall
that $\hat{{\cal G}}^E$ is just the quantization on the space of
functions over $(\overline{\cal A}\times\overline{\cal
U})^{\lambda}$, and that it has to be extended to our full reduced
Hilbert space of functions on $(\overline{\cal
A}\times\overline{\cal U})^{[\lambda]}$ by gauge covariance.

We can change the partial gauge fixing by performing a global
$SU(2)$-gauge transformation with $g\in
SU(2)\backslash\lambda(F)$. This transformation will conjugate the
holonomies appearing in the vector fields.  Using the formula
$$    \pd{g^{-1}hg}=g^T\pd{h}(g^{-1})^T  $$
which can be proved by using the chain rule, we get the
transformation rule
\begin{eqnarray*}
X_L^i(g^{-1}hg)&=&\tr\left((g^{-1}hg\tau_i)^T\pd{(g^{-1}hg)}\right)\\
&=&
  \tr\left((hg\tau_ig^{-1})^T\pd{h}\right)=\Ad_{ij}(g)X_L^j(h)
\end{eqnarray*}
where $\Ad_{ij}(g)$ are matrix elements in the adjoint
representation defined by
$\Ad_g\tau_i=g\tau_ig^{-1}=:\Ad_{ij}(g)\tau_j$. The right
invariant vector fields transform analogously. The transformed
Gau\ss\ constraint in the $g\tau_3g^{-1}$-gauge is now
\begin{eqnarray*}
  \hat{{\cal G}}_3^{\prime}[\Lambda] & = & i\hbar
    \sum_{v\in B}\Lambda(v)\left(\sum_{e(1)=v}
    X_L^3(g^{-1}h_eg)\right. -\\  &-& \left. \sum_{e(0)=v}X_R^3(g^{-1}h_eg)
    +X_L^3(g^{-1}h_vg)- X_R^3(g^{-1}h_vg)\right)\\
 & = & \Ad_{3i}(g)\hat{{\cal G}}_i[\Lambda]=:
    \hat{{\cal G}}[\Lambda_i].
\end{eqnarray*}
Here ${\cal G}_i$ are the components of the full $SU(2)$-Gau\ss\
constraint smeared with a function $\Lambda$, whereas ${\cal G}$
is the full Gau\ss\ constraint smeared with a function
$\Lambda_i:=\Ad_{3i}(g)\Lambda$. Allowing arbitrary $SU(2)$-gauge
transformations $g$, we can change $\Lambda_i$ arbitrarily. This
shows that the Gau\ss\ constraint on $\Phi_B$ is the full
$SU(2)$-constraint forcing $SU(2)$-spin networks with Higgs to be
gauge invariant.

We want to stress here the necessity of the mathematical apparatus
developed in Section~\ref{s:SymmState}: It enabled us to take the
role of the reduced gauge group into account, it provides us with
an interpretation of symmetric states as generalized states of the
unreduced theory and Theorem~\ref{ident} shows that all symmetric
states can be obtained by using the one-dimensional spin networks
used in the present subsection.
\subsubsection{Quantization of the Diffeomorphism Constraint}
We  conclude by making a few remarks on the diffeomorphism
constraint. It can be solved by group averaging where the
diffeomorphism group acts by dragging the Higgs vertices. But we
can alternatively regularize the diffeomorphism constraint and
solve it infinitesimally. The no-go theorem of Ref.\
\cite[Appendix C]{ALMMT} is evaded by the one-dimensional nature
of our graphs: Diffeomorphisms act only by dragging the ends of
edges and the Higgs vertices attached to them. But they can not
deform an edge transversally, and a longitudinal deformation can
always be absorbed by a reparameterization of the edge. The
quantized constraint is given by Lie derivatives on the Higgs
vertex positions leading to the well known solutions in
$\Phi_B^{\prime}$.

\section{Conclusions}

The main part of this paper is intended to define a space of
symmetric states in a diffeomorphism invariant theory of
connections, to investigate its properties and, especially, to
equip it with a calculus. This is achieved in Theorem~\ref{ident}.
The results show that there are some subtleties, mainly due to the
classical partial gauge fixing, which should be relaxed in the
quantum theory.

In order to illustrate different points of the general framework
we discussed several concrete examples:

 $2+1$ dimensional
gravity and spherically symmetric electromagnetism are quite easy
to deal with. Quantum symmetry reduction of these models yields
the expected results. Holonomy variables turned out to be well
suited in order to quantize the electromagnetic model and its
observables. Furthermore, the Ashtekar-Lewandowski measure
incorporates the classical reality conditions correctly. The
electromagnetic model exhibits a simple reduction of the degrees
of freedom to finitely many ones corresponding to the classical
theory.

Spherically symmetric gravity is the only example which has
necessarily a Higgs field in physically meaningful applications.
This property allows a non-trivial action of constraints. The
Gau\ss\ and diffeomorphism constraint are as easy to deal with as
in the full $3+1$ dimensional theory. Unfortunately, the
Hamiltonian constraint does not seem  to be more easily solvable
within the framework of loop quantum gravity with its spin network
states than in the full theory. In Refs.\ \cite{thika,kathi} it
was solved, but it, together with the diffeomorphism constraint,
had to be rephrased into a form not well suited for loop
quantization.

But regularized analogously to Ref.\ \cite{QSDI} the Hamiltonian
constraint operator might be defined and analyzed more easily,
because of the simplicity of one-dimensional graphs it acts on.
There is no place for it to create new edges and it will create
only Higgs vertices in the neighborhood of a Higgs vertex it acts
on, thereby changing the spin of the edge connecting the old and
the new Higgs vertex.  Note in this context that the appearance of
``Simon's subgraph'' of Ref.\ \cite{LM:Vertsm} is generic in the
symmetry reduced theory: The newly created vertices can always be
shifted to their neighboring vertices. Details and further
developments  are left for forthcoming publications \cite{bo3}.

A first physical application of the above considerations in the
case of gravity is to calculate the spectrum of the area operator
in the spherically symmetric sector \cite{bo4}.

\section*{Acknowledgements}

We thank T.\ Strobl for discussions and for bringing Ref.\
\cite{Brodbeck} to our attention.

M.~B.\ wants to thank the Max-Planck-Institut f\"ur
Gravitationsphysik, Potsdam, for its hospitality, Professor H.\
Nicolai for the invitation and, especially, T.~Thiemann for
proposing it and for stimulating discussions and suggestions. He
thanks the DFG-Graduierten-Kolleg ``Starke and elektroschwache
Wechselwirkungen bei hohen Energien'' for a PhD fellowship.

\end{document}